\documentclass[lettersize,journal]{IEEEtran}
\usepackage{amsmath,amsfonts}
\usepackage{algorithmic}
\usepackage{algorithm}
\usepackage{array}
\usepackage[caption=false,font=normalsize,labelfont=sf,textfont=sf]{subfig}
\usepackage{textcomp}
\usepackage{stfloats}
\usepackage{url}
\usepackage{verbatim}
\usepackage{graphicx}
\usepackage{cite}
\begin{document}

\title{Multi-Functional Metasurfaces with M-Type Ferrites: Shaping the Future of mmWave Absorption and Beam Steering}

\author{Nohgyeom~Ha,
	Horim~Lee,
	Min~Jang, 
	Gyoungdeuk~Kim,
	Hoyong~Kim, 
	Byeongjin~Park,
	Manos~M.~Tentzeris,~\IEEEmembership{Fellow,~IEEE,}   
	and Sangkil~Kim,~\IEEEmembership{Senior Member,~IEEE}
\thanks{This research was supported by National R\&D Program through the National Research Foundation of Korea (NRF) funded by Ministry of Science and ICT (RS-2025-02304267). (\textit{Corresponding author: Sangkil Kim})}
\thanks{Nohgyeom Ha, Gyoungdeuk Kim, Hoyong Kim, and Sangkil Kim are with the Department of Electronics Engineering, Pusan National University, Busan 46241, South Korea.}
\thanks{Horim~Lee, Min~Jang, and Byeongjin Park are with the Composites \& Convergence Materials Research Division, Changwon, 51508, South Korea.}
\thanks{Manos M. Tentzeris is with the School of Electrical and Computer Engineering, Georgia Institute of Technology, Atlanta, GA 30332 USA. (e-mail: ksangkil3@pusan.ac.kr; etentze@ece.gatech.edu)}}

\markboth{Journal of \LaTeX\ Class Files,~Vol.~14, No.~8, August~2021}%
{Shell \MakeLowercase{\textit{et al.}}: A Sample Article Using IEEEtran.cls for IEEE Journals}


\maketitle

\begin{abstract}
This paper presents a comprehensive review/tutorial on multi-functional metasurfaces integrated with M-type ferrite materials for millimeter-wave (mmWave) absorption and beam control. 
As wireless communication systems transition toward beyond-5G architectures, including non-terrestrial networks (NTNs), the demand for adaptive, low-profile electromagnetic surfaces that can manage interference while enabling beam reconfiguration becomes increasingly critical. 
Conventional metasurfaces often struggle to simultaneously achieve high absorption and beamforming over wide frequency ranges due to intrinsic material and structural limitations. 
This paper reviews the state-of-the-art in metasurface design for dual-functionality—particularly those combining frequency-selective magnetic materials with periodic surface lattices—to enable passive, compact, and reconfigurable reflectors and absorbers. 
Special emphasis is placed on the role of M-type ferrites in enhancing absorption via ferromagnetic resonance, and on the use of surface-wave trapping mechanisms to achieve narrowband and broadband functionality. 
A case study of a ferrite-based hybrid “reflectsorber” (reflectorarray + absorber) is presented to demonstrate key design concepts, analytical models, and application scenarios relevant to satellite, UAV, and NTN ground station deployments. Future directions for low-loss, tunable, and scalable metasurfaces in next-generation wireless infrastructures are also discussed.
\end{abstract}


\section{Introduction}

\begin{figure}
\centering
\includegraphics[width=1\linewidth]{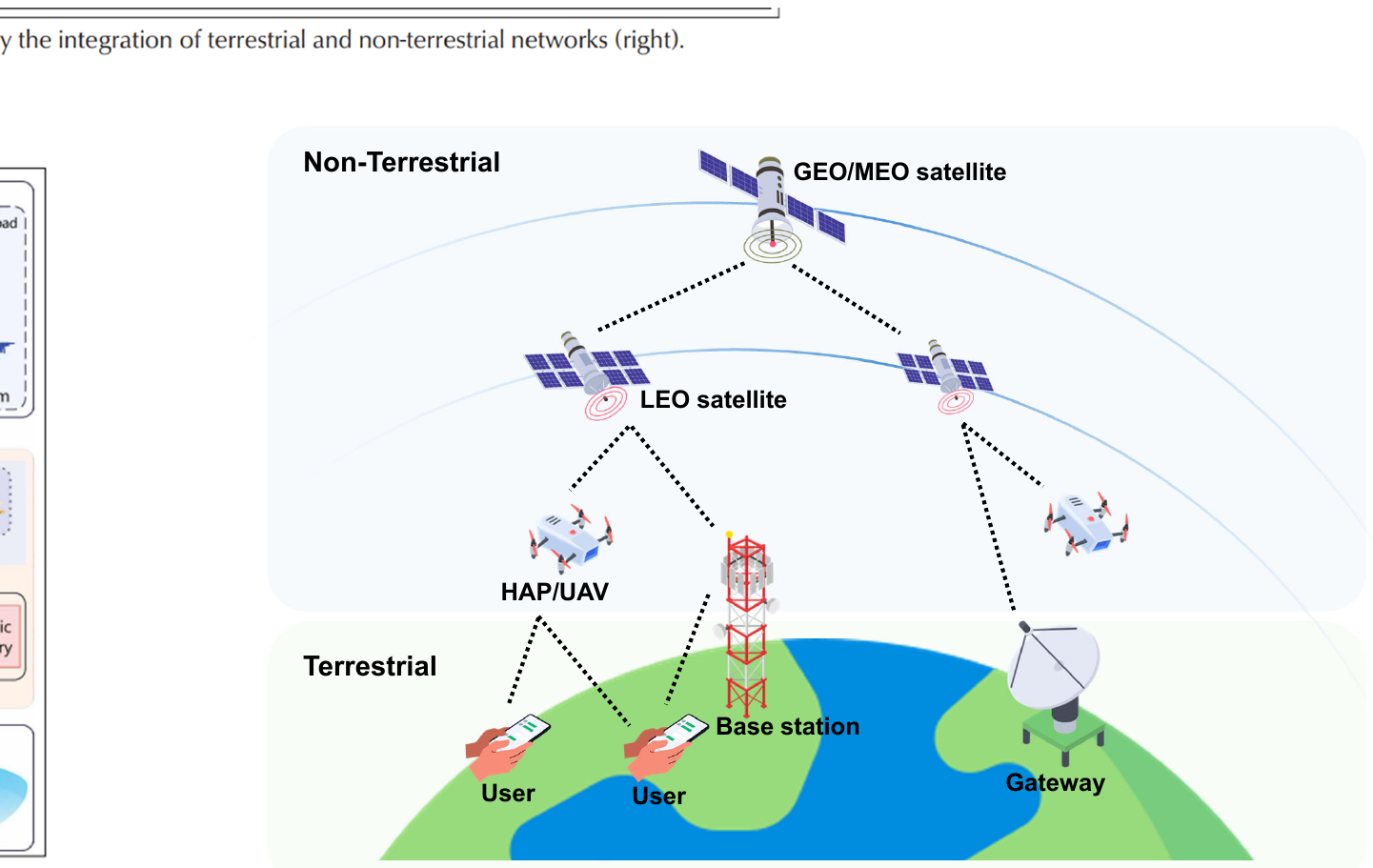}
\caption{Non-terrestrial network architecture involving LEO satellites and HAPS.}
\label{fig:Non-terrestrial_network}
\end{figure}

The expansion of non-terrestrial networks (NTNs) including high-altitude platform systems (HAPS) and low earth orbit (LEO) satellite constellations is drastically increasing the density of transmitters and receivers in the wireless ecosystem, as shown in Fig. \ref{fig:Non-terrestrial_network} \cite{su2019broadband, geraci2022integrating, kodheli2020satellite, giordani2020non, kurt2021vision, alfattani2023resource, tekbiyik2021channel}.
While these emerging platforms promise global coverage and low-latency links, they often share spectrum with legacy systems, leading to escalating interference challenges. 
For example, LEO mega-constellations plan to operate in Ku/Ka/V bands that overlap with existing geostationary (GEO) satellite and terrestrial cellular allocations, resulting in severe co-channel interference if unmitigated \cite{su2019broadband, geraci2022integrating, kodheli2020satellite}.
Likewise, HAPS platforms, which provide connectivity from the stratosphere, intend to use licensed sub-6 GHz mobile bands for a scenario that demands careful coordination with ground networks to prevent cross-tier interference \cite{kurt2021vision, alfattani2023resource, tekbiyik2021channel}. 
In general, modern hybrid networks frequently operate on closely adjacent frequency spectra, and conventional broadband approaches struggle with such coexistence issues. 
These trends make interference mitigation a critical concern in beyond-5G communications.

One promising approach to address these challenges is the use of metasurface-assisted communications.
Reconfigurable intelligent surfaces (RIS), or programmable metasurfaces, enable dynamic control of electromagnetic wave propagation, surpassing the capabilities of traditional antennas and fixed filters \cite{cheng2022reconfigurable}.
Essentially, an RIS is a two-dimensional metamaterial array with tunable elements, enabling software-defined control over the phase, amplitude, polarization, and even frequency response of impinging signals. 
By deploying RIS panels on building facades, indoor walls, or aerial platforms, the wireless channel itself becomes controllable. 
This capability allows the network to direct and focus coverage or nullify undesired paths, thereby improving signal quality and reducing interference. 
In fact, intelligent metasurfaces have been proposed to actively manipulate the propagation environment to mitigate interference from nearby cells or other unwanted transmitters. 
Compared to static filtering or coordination alone, metasurface-assisted beamforming can boost desired link gains while simultaneously deflecting or attenuating signals that would otherwise cause inter-system interference. 
This makes RIS technology a natural complement in hybrid NTN deployments, offering adaptive beam steering and spatial isolation without the need for power-hungry active relays.
However, active metasurfaces generally require an additional microcontrollers and power supplies to control their operating states, which will incur significant size, weight, and power.

\begin{figure}
\centering
\includegraphics[width=1\linewidth]{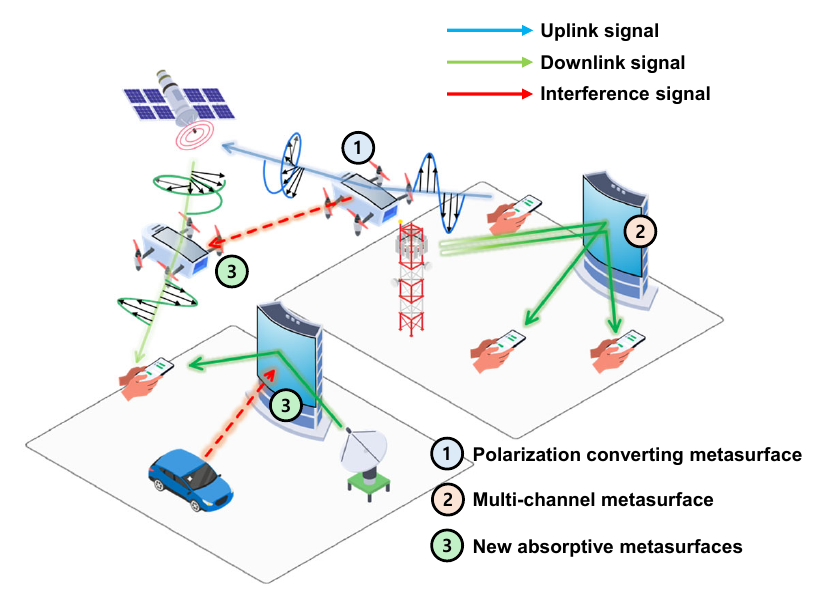}
\caption{Multi-functional metasurface involved in the next-generation communication with non-terrestrial networks, improving overall communication performances}
\label{fig:Multi-functional_metasurfaces}
\end{figure}
 
Passive multi-functional metasurfaces can be alternatives in developing a low-cost, compact, and power-independent ground station for NTNs with pre-defined orbits or locations.
Two prime examples of these multi-functionalities, and the focus of this work, are multi-band polarization converting and multi-channel spin-decoupled metasurfaces \cite{xu2018completely, ding2019dual, wu2021wideband, zhu20243, akram2020ultrathin, zhuang2018design, guo2019multi, ahmed2021multifunctional, wu2017broadband, ghosh2022low, ahmed2024multifunctional, rauf2025experimental}. 
The multi-functionality, such as a dual-band operation, is necessary for the metasurface to handle separate frequency bands of uplink and downlink, ensuring reliable communication between ground stations and LEO satellites.
Also, the metasurface with the multi-channel response can be utilized for a demanding task improving throughputs, such as adaptively splitting EM signals from ground stations to moving LEO satellites.
Numerous studies have highlighted the substantial benefits of metasurfaces \cite{pan2023pmsat, wang2024gpms, qian2022millimirror}.

Beyond simply reflecting or transmitting signals toward intended receivers, next-generation multi-functional metasurfaces are evolving into interference-free platforms to cope with complex interference scenarios. 
In particular, there is a strong rationale to integrate absorptive and beam-steering functionalities into a single metasurface architecture. 
Traditional reflecting surfaces lack any mechanism to damp unwanted transmissions; however, by introducing an absorption mechanism in selected elements or frequency bands, a metasurface can actively dissipate interfering energy before it propagates further. 
Researchers have begun to demonstrate such dual-purpose designs – for instance, metasurface arrays that can switch between a reflective beamforming mode and an transmitting mode to effectively “wall off” certain signals \cite{phon2020dynamically}.
The ability to selectively absorb stray or cross-polarized signals for undesired frequencies is especially valuable in dense NTN environments, where overlapping coverage areas and multipath can generate strong interference. 
By uniting absorption with rerouting (reflection/refraction) within different frequency bands, a single metasurface can thus serve as both a smart reflector for wanted beams and a spatial filter for unwanted ones.

This paper introduces, for the first time, a novel structure that combines a metasurface with an M-type ferrite slab, designed for advanced communication system with NTNs. 
In a higher-frequency region, high-quality factor resonances generated in periodic structures are analyzed by using the Floquet expansion method based on the dielectric substrate with the surface lattice.
While the resonance frequencies of the periodic structure can be tailored through dimensional control, the inherently high-Q factor must be suppressed to enable broadband functionalities such as filtering or absorption. 
Conventional dielectric substrates pose limitations in increasing dielectric losses, which constrains their applicability in broadband designs. 
To address this, an M-type ferrite material is employed to effectively dissipate the trapped electromagnetic energy in the periodic structure. 
This magnetic material offers frequency-dispersive properties, allowing independent control of spectral behavior across different frequency bands.
The proposed design can be applied in integrating the absorptive property and any beam-shaping function. 
This paper presents a concept and operational theory of the `reflectsorber'–a hybrid device that functions both as a reflectarray and an absorber for practical implementations. 
The validity of the theoretical model of the reflectsorber is demonstrated through the design and fabrication of the metasurface, which operates as a reflectarray at 38 GHz and an absorber at 77 GHz.

\section{Multi-functional Metasurfaces: Review}

Metasurfaces are ultrathin, engineered 2D materials composed of sub-wavelength elements that can manipulate electromagnetic waves in extraordinary ways, far beyond the limits of natural materials. 
By judicious design of these planar meta-atoms, one can finely tailor the phase, amplitude, and polarization of reflected or transmitted waves. 
In the microwave and millimeter-wave (mmWave) bands, such metasurfaces have enabled novel flat optics and antennas – including anomalous reflection/refraction \cite{diaz2017generalized, wong2018perfect, movahediqomi2023comparison, vuyyuru2023efficient}, invisibility \cite{moreno2018wideband, sima2018combining, lv2020hybrid}, and vortex beam generation \cite{huang2019high, li20203d, huang2021generation} – all in a compact form factor. 
This powerful wavefront control comes with significant benefits in high-frequency communication systems where size, weight, and power (SWaP) are at a premium.
Meanwhile, the metasurfaces with a single function need to be developed for satisfying next-generation communication networks.




The motivation for multi-functional metasurfaces arises when a single engineered surface is designed to perform several electromagnetic functions at once, further reducing component count, size, and power consumption. 
For example, a single metasurface panel might simultaneously act as a high-gain antenna and a radar cross section (RCS) reductor, obviating the need for separate antenna feeds and RCS surfaces \cite{fan2019low}.
In satellite systems, this means one compact aperture can handle dual-polarized or circularly polarized links without the weight of mechanical polarizers, while in terrestrial 5G systems a metasurface can combine beamforming with interference suppression in one device.
Such integration yields clear SWaP benefits: fewer physical parts, lighter payloads, and no lossy adapters, all translating to improved efficiency. 
In addition, multi-function metasurfaces can improve performance and flexibility, offering capabilities like dynamic polarization multiplexing or adaptive absorption that legacy hardware antennas cannot readily achieve.

Two prime examples of these multi-functionalities, and the focus of this work, are multi-band polarization converting and multi-channel spin-decoupled metasurfaces \cite{xu2018completely, ding2019dual, wu2021wideband, zhu20243, akram2020ultrathin, zhuang2018design, guo2019multi, ahmed2021multifunctional, wu2017broadband, ghosh2022low, ahmed2024multifunctional, rauf2025experimental}. 
Metasurfaces can serve as efficient polarization converters, transforming an incident wave's polarization state (for instance, from linear to circular polarization) over broad bandwidths, as shown in Fig. \ref{fig:Multi-functional_metasurfaces}. 
This is extremely valuable in satellite communications, where links often require circular polarization to mitigate Faraday rotation and polarization misalignment. 
A metasurface polarization converter can replace bulky quarter-wave plates or orthomode transducers, achieving the desired polarization in a thin planar layer. Likewise, metasurfaces enable dual-polarized reflectarrays in which a single aperture independently controls two orthogonal polarizations, as shown in Fig. \ref{fig:Multi-functional_metasurfaces}. 
Multi-channel metasurfaces, such as a dual-feed or dual-polarized metasurface, can generate two separate beams or communication channels on the same physical antenna, increasing the data throughput via polarization multiplexing. 
Compared to conventional single-polarized antennas, these metasurface-based solutions offer greater functionality without additional hardware, which is a decisive advantage in both satellite payloads and 5G base stations. 
In the following, we survey recent developments in these areas from polarization-converting metasurfaces to multi-channel metasurface designs, and also discuss a third emerging function: absorptive metasurfaces for interference management in hybrid terrestrial-satellite networks.



\begin{table*}[t]
\caption{Operation Comparison for Multi-functional Metasurfaces}
\label{Comparison_table}
\renewcommand{\arraystretch}{1.5}
\centering
\begin{tabular}{|c|c|c|c|}
\hline
\textbf{Related Work} & \textbf{Material} & \textbf{Beam-shaping Function} & \textbf{Interference Mitigation} \\ \hline
\cite{wu2021wideband} & F4B (Dielectric material) & O (Dual-LP channel reflectarray) & X   \\ \hline
\cite{ding2019dual} & Taconic TRF-43 (Dielectric material) & O (Dual-CP channel reflectarray & X   \\ \hline
\cite{guo2019multi} & FR4 (Dielectric material) & O (Dual-band dual-channel reflectarray) & X   \\ \hline
\cite{ahmed2021multifunctional} & FR4 (Dielectric material) & O (Multi-band polarization conversion) & X   \\ \hline
\cite{su2022ultrawideband} & F4B (Dielectric material) & X & O (Ultra-wideband RCS reduction)   \\ \hline
\textbf{This work} & M-type ferrite (Frequency selective material) & O (Reflectarray or Anomalous reflection) & O (Absorption)   \\ \hline
\end{tabular}
\end{table*}


\subsection{Multi-channel Metasurfaces With Spin-decoupled Elements}

One important class of multi-functional metasurface technology is the development of dual-polarized (or dual-feed) reflector whose elements can independently control orthogonal polarizations \cite{xu2018completely, ding2019dual, wu2021wideband, zhu20243, akram2020ultrathin, zhuang2018design, guo2019multi}. 
In a traditional reflectarray or phased array, managing two polarizations often requires duplicating hardware (e.g. separate arrays or interleaved elements for horizontal and vertical polarization) or using multi-layer structures with polarization-selective surfaces. 
The multi-functional metasurfaces offer a smarter solution: by designing each meta-element with anisotropic or multi-resonant features, one can impart a distinct phase shift to each polarization of the incident wave. 
In effect, the reflective metasurface can impose two independent phase profiles – one for, say, the $x$-polarized component and one for the $y$-polarized component – enabling full decoupled control of dual polarizations from the same aperture. 
This capability underpins a new class of dual-feed reflectarrays where two feed sources (often orthogonally polarized) illuminate the same metasurface, and the metasurface generates two different beams, one per polarization. 
The feeds could be, for example, a dual-polarized horn or two separate antennas, launching horizontal and vertical polarizations that are handled separately by the metasurface. 
Because the metasurface elements introduce low cross-coupling, each polarization can be beamformed or redirected independently, realizing beam multiplexing and polarization diversity with a single flat panel.

The benefits of such dual-polarized metasurface reflectarrays are substantial. 
First, they effectively double the communication capacity without doubling aperture size – two data streams can be sent/received on the same frequency band via orthogonal polarizations (polarization multiplexing). 
This is particularly advantageous for satellite communications and wireless backhaul links, where spectrum is precious and dual-polarization can roughly double the throughput. 
Indeed, a dual circularly polarized (dual-CP) reflectarray can transmit left-handed circularly polarized (LHCP) and right-handed circularly polarized (RHCP) beams on separate channels, greatly improving data rate and reducing polarization losses in a satellite link. 
Zhang et al. recently reported a Ku-band dual-CP reflectarray that radiates two independent beams for LHCP and RHCP at different angles ($\pm$20$^\circ$ from broadside) with high gain ~29 dB for each beam. 
Each unit cell in their design provides two phase responses such that the metasurface focuses the left-handed circularly polarized wave in one direction and the right-handed wave in another. 
Notably, they achieved this multiplexing in a single-layer metasurface by combining the dynamic phase (tuning element size to delay one polarization) and Berry phase (rotating elements to redirect the other polarization) within each cell. 
This approach circumvented the need for stacked polarizing layers and achieved broad 37\% bandwidth for both CP channels. The result is an ultrathin dual-CP antenna that provides two high-gain beams from one aperture, exemplifying the beam-multiplexing power of spin-decoupled metasurfaces.

A reflective coding metasurface reported in \cite{ding2019dual} can independently tune the phase functions for CP wavefronts with two orthogonal helicities.
The meta-atom in the metasurface has a low cross talk for linearly orthogonal polarized waves, which helps to introduce multi-OAM modes.
It shows that a incident CP wave is able to be split into two different OAM beams at a microwave region of 16 GHz.
Another multi-channel metasurface provides quadruplex channels for reflection and transmission modes, which is based on the meta-atom with linear polarized patches and phase delay lines \cite{zhu20243}.
Under LHCP incident wave, the metasurface can re-radiate the impinging wave into the four channels with transmitting space of RHCP/LHCP channels and reflecting space of RHCP/LHCP channel, covering 28--30 GHz.
Besides, the multimaterial inside the multi-channel metasurface was fabricated by utilizing the 3-D printing technique.

\subsection{Multi-band Metasurfaces for Polarization Transformations}

Another frontier of multi-functional metasurfaces enables polarization conversion, wherein an incoming wave of one polarization is re-emitted in another polarization state \cite{ahmed2021multifunctional, wu2017broadband, ghosh2022low, ahmed2024multifunctional, rauf2025experimental}. 
This functionality is critical in wireless systems that must accommodate different polarization schemes for compatibility or performance reasons. 
For example, satellite transponders often use CP to avoid polarization alignment issues and to support dual orthogonal channels (RHCP/LHCP), even though ground terminals or feeder links may transmit in linear polarization (LP). 
A metasurface polarization converter can bridge this gap by converting linear-to-circular polarization over the satellite downlink/uplink bands. 
Similarly, in terrestrial links and radar, converting one linear polarization to its orthogonal counterpart (linear-to-linear, or LP–LP conversion) can improve link reliability and reduce interference by aligning with the receiver’s preferred polarization or by exploiting polarization diversity. 
Metasurface-based converters achieve this without bulky microwave components, using sub-wavelength resonators to impart the required phase shifts to the incident waves.

As reported in \cite{ahmed2021multifunctional}, the research proposes a multifunctional metasurface designed for efficient polarization manipulation, capable of both cross- and circular-polarization conversion. 
Specifically, it provides linear-to-linear polarization conversion over the frequency bands of 5.3--5.4 GHz, 7.2–-8 GHz, and 12.3--13.76 GHz, while effectively converting linear polarization to circular polarization across broader ranges of 5.1--5.2 GHz, 5.6--6.85 GHz, 8.8--11.2 GHz, and 14.9--20.2 GHz. 
Due to its compact size, excellent angular stability up to 75$^\circ$, and multifunctional capability, the metasurface is highly promising for diverse applications within strategic C-, X-, Ku-, and K-band communications.
In \cite{wu2017broadband}, the other proposed metasurface with an active tunability shows the multi-functional polarization conversion with ultrabroad bandwidth and large angular tolerance. 
Especially, the microfluidic network was involved in every galinstan-based metasurface element and controled their orthogonal resonant modes continuously. 
As proof of concept, the multifunctional polarization converter was validated for linear-to-linear, linear-to-circular, and linear-to-elliptical polarization conversions by controlling the galinstan within the microfluidic channels. 
The possibility of tuning the incident EM wave to any desired polarization state was suggested, showing intriguing new opportunity to create active devices with dynamic polarization control.

\begin{figure*}
\centering
\includegraphics[width=1\linewidth]{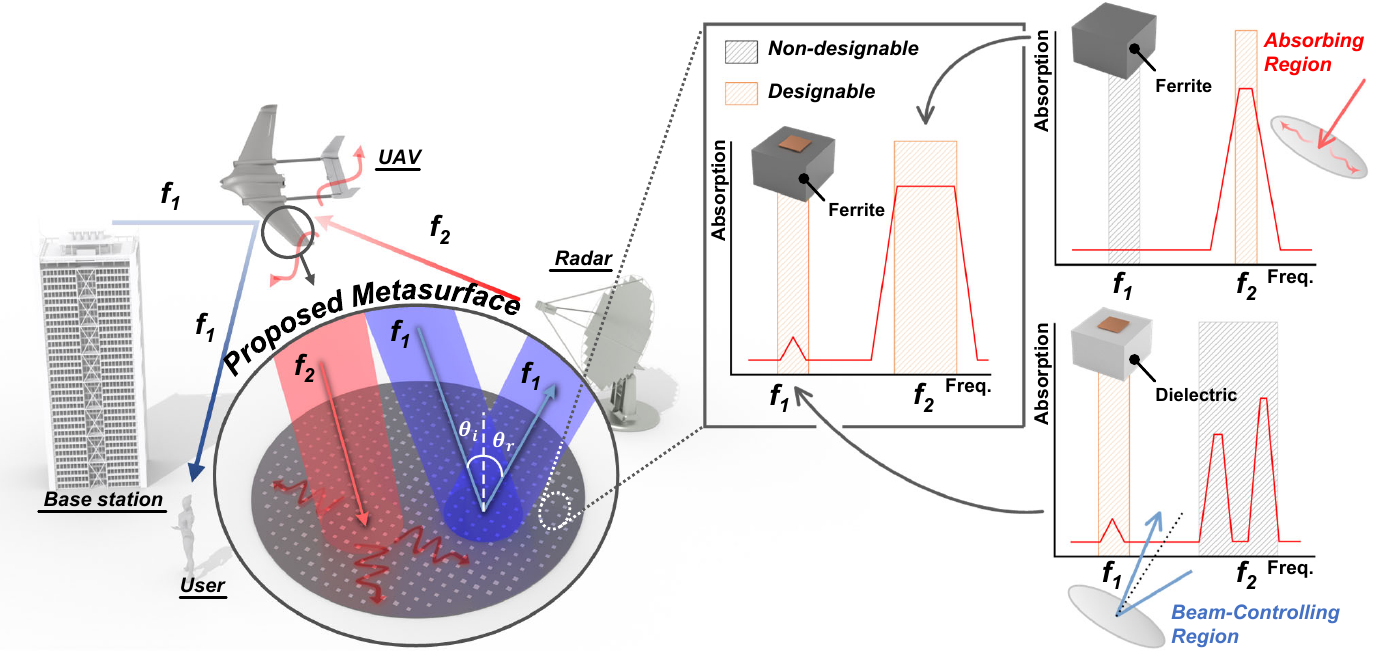}
\caption{Schematic of the proposed metasurface with the application sketch. The inset of the schematic shows the different operation of the unit-cell of the metasurface based on M-type ferrite and conventional dielectric materials.}
\label{fig:Application_description}
\end{figure*}

\subsection{Interference Management via Absorptive Metasurfaces in Hybrid Networks}

As communication networks become increasingly heterogeneous – integrating terrestrial 5G/6G cells with non-terrestrial nodes like satellites and high-altitude platforms – interference management emerges as a crucial challenge \cite{azari2022evolution, mismar2024uncoordinated, rinaldi2020non}.
These hybrid networks involve a mix of signals at various frequencies and propagation directions, raising the risk of unwanted interactions: for instance, a satellite downlink might interfere with terrestrial base station receivers, or reflections off buildings might desensitize a nearby radar. 
Multi-functional metasurfaces can play a transformative role here by incorporating absorptive or scattering-reducing features to mitigate interference signals while still performing their primary communication functions. 
In essence, an absorptive metasurface can act as a selective shield – allowing or directing desired signals while simultaneously suppressing or diffusing stray and potentially interfering signals.

One application of absorptive metasurfaces is in controlling the radar cross-section (RCS) or general reflectivity of communication components \cite{su2022ultrawideband, leung2021broadband, zhao2016broadband, xu2022low, zhou2020absorptive, caizzone2023spatial, song2024origami}.
A conventional antenna or reflecting surface can produce strong incidental reflections (e.g., backscatter towards a radar or echoes into other communication channels) that contribute to electromagnetic interference. 
By engineering a metasurface to be bifunctional – that is, to both steer the main beam and absorb energy outside the main beam or band – one can substantially reduce these undesired reflections. 
For example, an absorptive metasurface of \cite{su2022ultrawideband} designed an absorptive coding metasurface for ultra-wideband RCS reduction that uses a resistive layer to absorb much of the incident energy while employing a 1-bit phase coding on the reflected portion to scatter the rest in diverse directions. 
Their metasurface's unit cell was an anisotropic resistive patch that acted as a lossy polarization converter – co-polarized reflections were kept below –10 dB from 1.5 to 40 GHz, with the remaining cross-polarized reflections randomized in phase to cancel out coherently. 
The practical implication is that a communication panel based on such a metasurface would be far less visible to detection and would scatter negligible energy into other systems' spectrum, thereby minimizing interference in a dense electromagnetic environment.

\begin{figure*}
\centering
\includegraphics[width=0.8\linewidth]{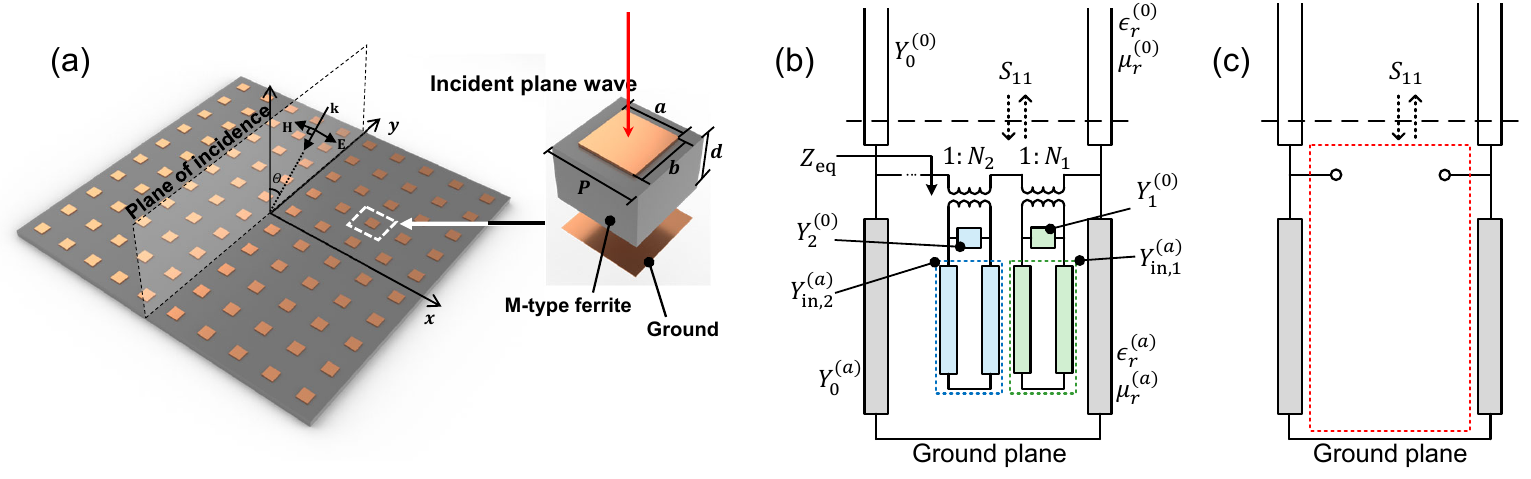}
\caption{The configuration of the metasurface based on the M-type ferrite with the transverse electric (TE) plane wave. (a) Scheme of the patch array of finite thickness under oblique incidence, (b) equivalent circuit network corresponding to the proposed structure, and (c) equivalent circuit network in case of the transverse resonance of any $h$-th harmonic.}
\label{fig:Configuration_metasurface}
\end{figure*}

In the context of hybrid terrestrial/non-terrestrial networks, absorptive metasurfaces can be strategically deployed to isolate the two layers of the network. 
Consider a building-mounted intelligent reflecting surface that assists mmWave 5G signals: if that surface is also exposed to satellite downlink signals (e.g., in Ka-band or Q-band), a portion of the satellite signal might otherwise reflect and interfere with ground users or create multi-path issues. 
By designing the metasurface with frequency-selective absorption, it can absorb the satellite band signals (or any other unwanted band) while still efficiently reflecting the 5G band signals towards their target \cite{caizzone2023spatial}.
This frequency-selective absorption can be achieved by incorporating resistive elements tuned to the interfering band, converting those unwanted waves into heat. Essentially, the metasurface becomes a spatial filter, providing gain where needed and loss where beneficial. 
The result is improved electromagnetic compatibility between coexisting systems without adding a complex techniques. 
Similarly, absorptive metasurfaces can be used on satellite surfaces or ground station radomes to reduce outgoing sidelobe energy and echoes that might interfere with other satellites or radio services \cite{song2024origami}.
Because metasurface absorbers are thin and can cover large areas, they lend themselves to coating structures like aircraft, high-altitude platforms, or base-station towers to suppress scattering and interference.

It should be noted that purely absorbing metasurfaces inherently dissipate power, which in a communication context means some signal loss. 
Thus, designers often seek a balance: absorptive metasurfaces in communications are typically engineered to absorb only unwanted energy (out-of-band signals, high-angle sidelobes, reflections back toward interferers) while preserving high efficiency for the desired in-band, in-beam signals.
It is challenging to realize an integrating metasurface with absorbing and beam-shaping properties, without a performance degradation in both absorbing and radiating properties.
A new design methodology is required, to realize the bi-functional metasurface within a single slab.
In this work, a novel metasurface architecture is proposed by integrating a frequency-selective material with a periodic structure, enabling multi-functionality.
The proposed metasurface simultaneously supports both absorption and beam-steering functionalities, ensuring interference-free communications for undesired frequency band, as Case 3 in Fig. \ref{fig:Application_description}.
Furthermore, additional features, such as polarization conversion and multi-channel operation, can be seamlessly incorporated into the design while maintaining its absorptive characteristics as shown in Table \ref{Comparison_table}.



\begin{figure}
\centering
\includegraphics[width=1\linewidth]{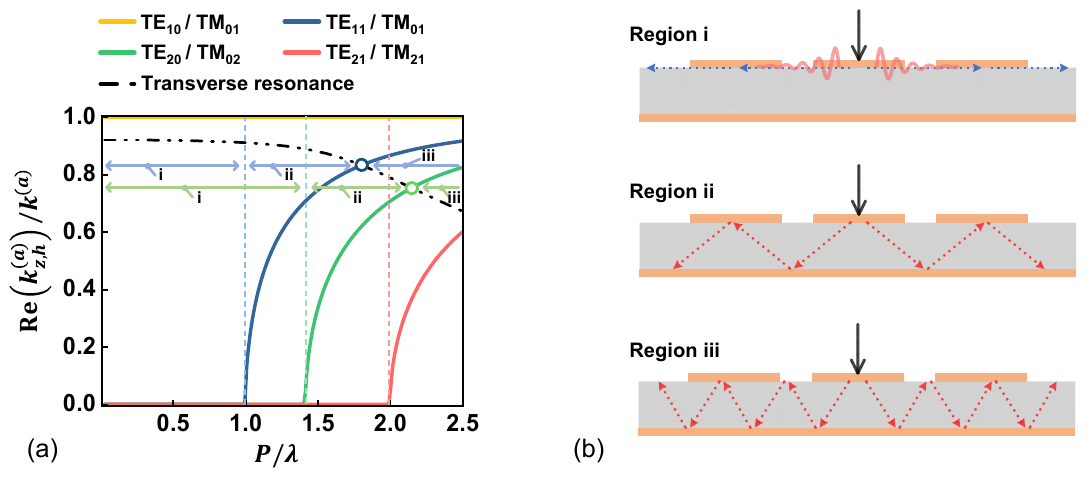}
\caption{(a) Real part of the normal wavenumber for the $h$-th harmonics with the fixed periodicity and (b) the description of the propagating properties for the operating regions.}
\label{fig:Dispersion_diagram}
\end{figure}

\section{Analysis of Surface Waves in Metasurfaces}

\subsection{Floquet Harmonic Surface Waves on Dielectric Substrates}

It is considered that the metasurface composed of zero-thickness metallic patches lies on either a dielectric or ferrite substrate. 
As shown in Fig. \ref{fig:Configuration_metasurface}(a), a TE-polarized plane wave is incident at an angle $\theta$ in the principal $yz$ plane of the patch array
The impinging field parallel to the periodic structure is in the $E$-plane, inducing the surface currents on the metasurface.
Reflection and transmission from the metasurface with periodic structure correspond to the interference of an infinite set of propagating and evanescent plane waves, i.e., Floquet harmonics.
As shown in Fig. \ref{fig:Configuration_metasurface}(b), the propagation of the incident and reflected plane waves can be modeled by the equivalent circuit network (ECN) by waveguide and the impedance matching concepts (see Appendix) \cite{rodriguez2015analytical, mesa2018unlocking}.
When the plane wave arrives at the metasurface with periodic structure, the scattered harmonics can be excited along the surface lattice.
The wave numbers ($k_{x,m}$ and $k_{y,n}$) in the $x$- and $y$-axes associated with the Floquet harmonics of order $m$ and $n$ are denoted by
\begin{align}
&k_{x,m} = \frac{2m\pi}{P} \\
&k_{y,n} = k_0 \sin{\theta} + \frac{2n\pi}{P}
\end{align}
where $m$ and $n$ are integer numbers ($m, n = 0,\pm1,\pm2,\cdots$), and $k_0$ is the free-space wavenumber.
Inside a substrate, the normal wave-number of $h$-th harmonic with a mode pair of $mn$ are defined as follows:
\begin{flalign}
&k^{(a)}_{z,h} = 
	\begin{cases}
	\beta_{h}^{(a)} = \sqrt{\{ k^{(a)} \}^2 - |\mathbf{k}_{\mathrm{t},h}|^2}, & k^{(a)} \ge |\mathbf{k}_{\mathrm{t},h}| \\
	\ -j \alpha_{h}^{(a)} = -j\sqrt{|\mathbf{k}_{\mathrm{t},h}|^2 - \{ k^{(a)} \}^2}, &  k^{(a)} < |\mathbf{k}_{\mathrm{t},h}|
	\end{cases}
\label{eq:wavenumber}
\end{flalign}
where the type of substrate is indicated by the superscript ($a$). 
The superscripts ($\mathrm{d}$) and ($\mathrm{m}$) for ($a$) indicate the dielectric substrate with $\mu_r = 1$ and the magnetic substrate with $\mu_r \ne 1$, respectively.
If the superscript is ($0$), the substrate is considered to be the free space with $\epsilon_r = 1$ and $\mu_r = 1$.
For a lossless material, the harmonics are represented as purely real or purely imaginary wavenumbers, which indicates whether it is a propagating or evanescent wave.

\begin{figure}
\centering
\includegraphics[width=0.7\linewidth]{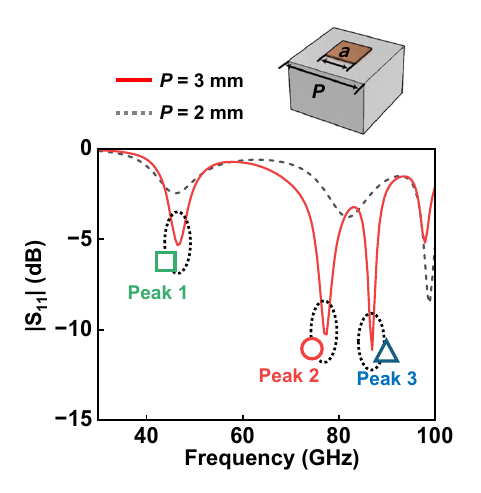}
\caption{Full-wave electromagnetic simulation for reflection coefficient of a metasurface with $P = 2$ and $P = 3$ mm}
\label{fig:Comparison_P_2mm_3mm}
\end{figure}

As shown in Fig. \ref{fig:Dispersion_diagram}(a), the normal component of the propagating modes impinging the substrate appears at a certain wavelength with respect to an angle of $\theta$ and a periodic of $P$.
The regions corresponding to each propagation property exactly depend on the periodicity: (i) evanescent, (ii) guided, and (iii) trapped modes \cite{munk2005frequency, garcia2012simplified}.
Fig. \ref{fig:Dispersion_diagram}(b) shows the surface wave of the $h$-th harmonic is exponentially decayed within frequencies of Region \textbf{i}.
In Region \textbf{ii}, the surface wave of the $h$-th harmonic starts to impinge the substrate, while guided between the metasurface and the ground plane.
The surface wave of the $h$-th harmonic flows along the metasurface when satisfying the transverse resonance condition that $Y_1^{(0)} + Y_\mathrm{in,1}^{(a)} = 0$.
The transverse resonances corresponding to TM$_{01}$, TM$_{11}$, and TE$_{10}$ nides are denoted as $f_\mathrm{tr1}$, $f_\mathrm{tr2}$, and $f_\mathrm{tr3}$, respectively.
As shown in Fig \ref{fig:Configuration_metasurface}(c), the dielectric substrate are only considered without the influence of the $h$-th harmonics because the input impedance ($Z_\mathrm{eq}$) is infinite at the transverse resonance.
After transverse resonance of the $h$-th harmonic, the guided waves of the $h$-th harmonics with a slower wavenumber are trapped within the surface lattice for Region \textbf{iii}.
For all the trapped waves of the $h$-th harmonic, interesting resonances with high-quality factor are generated by facing the incident wave.

\begin{figure}
\centering
\includegraphics[width=1\linewidth]{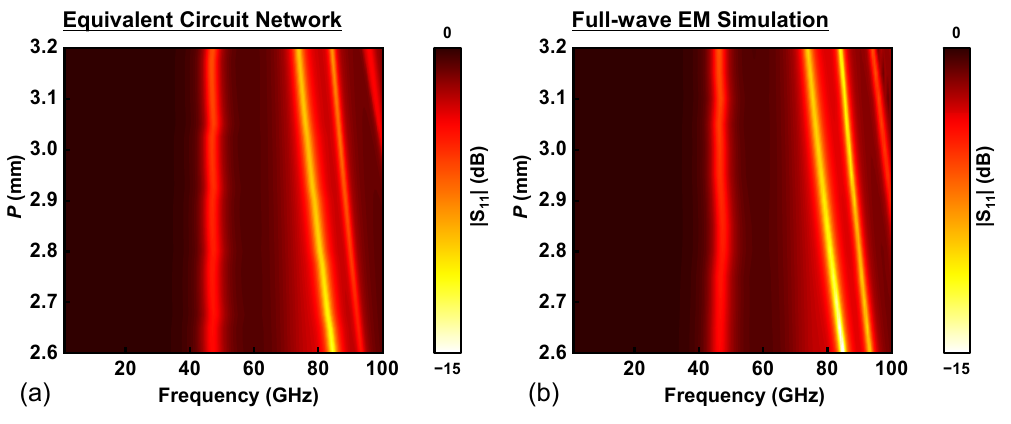}
\caption{Reflection cofficient of the metasurface with periodic structure as a function of the unit-cell spacing calculated by (a) 3D FEM simulation and (b) ECN model approach}
\label{fig:Dielectric_case_periodicity_HFSS_ECN}
\end{figure}

As illustrated in Fig. \ref{fig:Comparison_P_2mm_3mm}, the interesting influence of space harmonic resonance is examined by comparing the cases of periodicities $P = 2$ mm and $P = 3$ mm, where the metasurface with $a = 1$ mm is positioned on a dielectric substrate with a relative permittivity of $\epsilon_r = 6.5$.
The substrate thickness is optimally set to a quarter wavelength ($\lambda / 4$) in order to exploit the phase cancellation between the incident wave and the trapped wave of the $h$-th harmonic.
In the common case with a unit-cell spacing, the fundamental resonance of the metasurface with periodic structure can be easily expected as $f_r \approx 2c/(a\sqrt{\epsilon_r})$.
While the peak 1 is derived from the fundamental mode, an additional resonance modes emerge when the periodicity increases with a fixed size of $a$.
As shown in Fig. \ref{fig:Dispersion_diagram}, the extra trapped surface waves resonated within the periodic metasurface can be introduced for increasing the unit-cell spacing ($P$).
The quality factors of the new resonances have higher values of more than 100, which provides a potential of absorbing and filtering applications with high performance.

\begin{figure}
\centering
\includegraphics[width=0.7\linewidth]{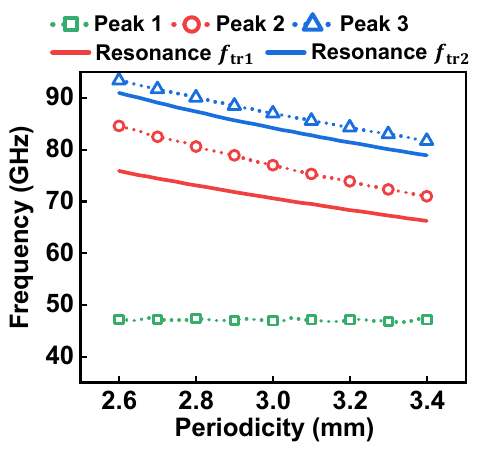}
\caption{Variation of peak points and transverse resonances in the reflection coefficient regarding to the unit-cell spacing ($P$).}
\label{fig:Peak_variation}
\end{figure}
\begin{figure}
\centering
\includegraphics[width=1\linewidth]{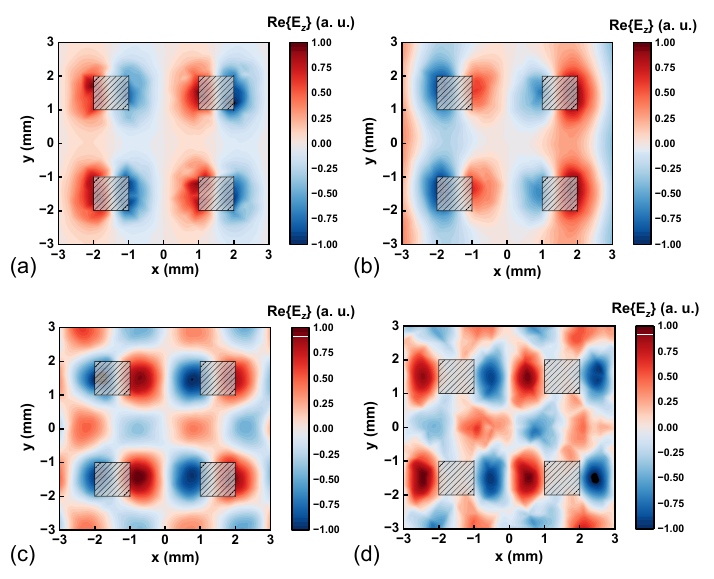}
\caption{Simulated $E_z$-field distributions at a few selected frequencies of (a) 48, (b) 60, (c) 70, and (d) 85 GHz. At frequencies of 48 and 60 GHz, the $E_z$-field distribution is concentrated around the periphery of each meta-unit. However, the $E_z$-field is spatially distributed within the surface lattice at 70 and 85 GHz, showing the effect of the trapped surface wave of the $h$-th harmonic.}
\label{fig:Dielectric_E_z_distribution}
\end{figure}

To further specify the trapped surface wave of these additional resonances, the analytic approach of ECN modeling is applied on this metausrface.
The influence of $h$-th space harmonic caused in a periodic structure can be simplified into a transmission line based on waveguide concept of ECN model.
For a fixed size of $a = 1$ mm and a varying spacing of $P$, the finite element method (FEM) simulation and the numerical calculation based on ECN modeling are conducted, while representing the excellent agreement as shown in Fig. \ref{fig:Dielectric_case_periodicity_HFSS_ECN}(a) and (b).
It is notable that the resonance frequencies are determined by the periodicity variation. 
Also, the comparison confirms that the periodic structure resonances are derived from the impedance matching between the incident wave and the trapped wave of the $h$-th harmonic.
As shown in Fig. \ref{fig:Peak_variation}, the peak 2 and 3 generated by the periodic structure resonance follow behind the transverse resonance.
The unit-cell spacing ($P$) can be managed for the frequencies of the peaks to be in an interesting region for obtaining the desired filtering or absorbing property.

\begin{figure}
\centering
\includegraphics[width=1\linewidth]{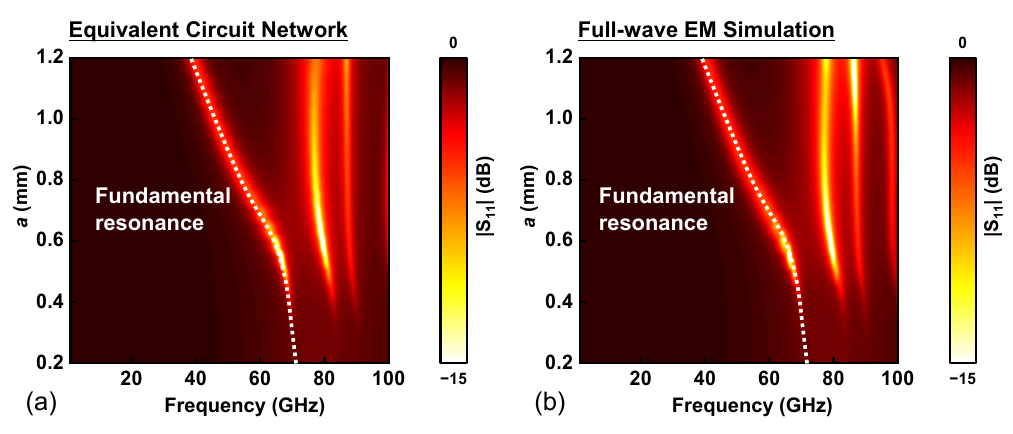}
\caption{Reflection cofficient of the periodic metasurface as a function of the square pattern size ($a$) calculated by (a) 3D FEM simulation and (b) ECN model approach}
\label{fig:Dielectric_case_square_size_HFSS_ECN}
\end{figure}
\begin{figure}
\centering
\includegraphics[width=0.7\linewidth]{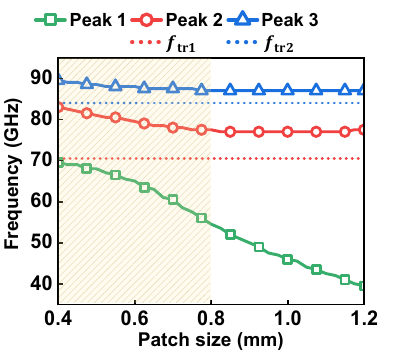}
\caption{Variation of peak points and transverse resonances in the reflection coefficient regarding to the square pattern size ($a$).}
\label{fig:Peak_variation_square}
\end{figure}

Fig. \ref{fig:Dielectric_E_z_distribution} represents distributions of the $z$-direction E-field ($E_z$) at the ground plane for a selected few frequencies.
As shown in Fig. \ref{fig:Dielectric_E_z_distribution}(a) and (b), the $E_z$-fields at 48 and 60 GHz demonstrate the fundamental resonances of the metasurface with periodic structure and its sustained resonance beyond the resonant frequency.
In these cases, the $E_z$-fields are most intense on the border of the square pattern.
As shown in Fig. \ref{fig:Dielectric_E_z_distribution}(c) and (d), a considerable amount of $E_z$-field is bound within the periodic structure far from the square pattern at 70 and 85 GHz.
The discrepency of the $E_z$-field distributions at 70 and 85 GHz is caued by the different propagating modes of the surface waves. 
These distributions prove the resonances of the higher frequencies are generated by the trapped waves propagating parallel to the plane of the surface.
However, the remaining $E_z$-field distribution adjacent to the patch means the absorption mechanism does not work with the conventional dielectric material.

\begin{figure}
\centering
\includegraphics[width=1\linewidth]{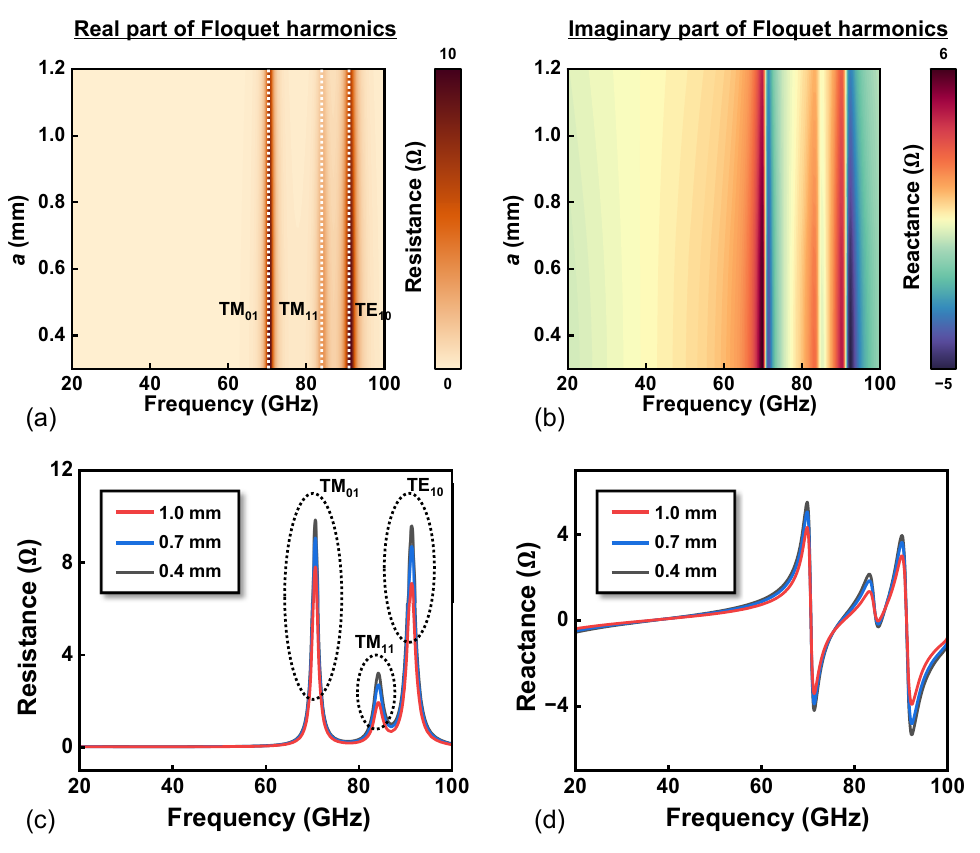}
\caption{(a) Resistance and (b) reactance in the input impedance of the Floquet harmonics regarding to the square pattern size ($a$) and the overall frequencies. (c) Resistance and (d) reactance components of the input impedance associated with the Floquet harmonics for certain values of the parameter $a$, which exhibit singularities that remain invariant under fixed periodic modulation}
\label{fig:Impedance_harmonics}
\end{figure}

For a fixed spacing of $P = 3$ mm, the square pattern size ($a$) serves as a determining parameter for the operating frequency of the fundamental resonance, as shown in Fig. \ref{fig:Dielectric_case_square_size_HFSS_ECN}(a) and (b).
Meanwhile, Fig. \ref{fig:Peak_variation_square} shows that the periodic structure resonances can be varied on the certain condition of $a$.
Due to the relatively small size of the pattern compared to the periodicity, higher-order harmonic modes within the periodic structure are challenging to excite with significant intensity. 
Instead, the fundamental resonance induced by the square patterns predominantly governs the surface lattice behavior.
The influences for the unit dimension of the metasurface are easily described by the impedance matching concept of the ECN model, which approximates the relative dominance of the eigenmodes into the impedance parameters.
\begin{figure}
\centering
\includegraphics[width=1\linewidth]{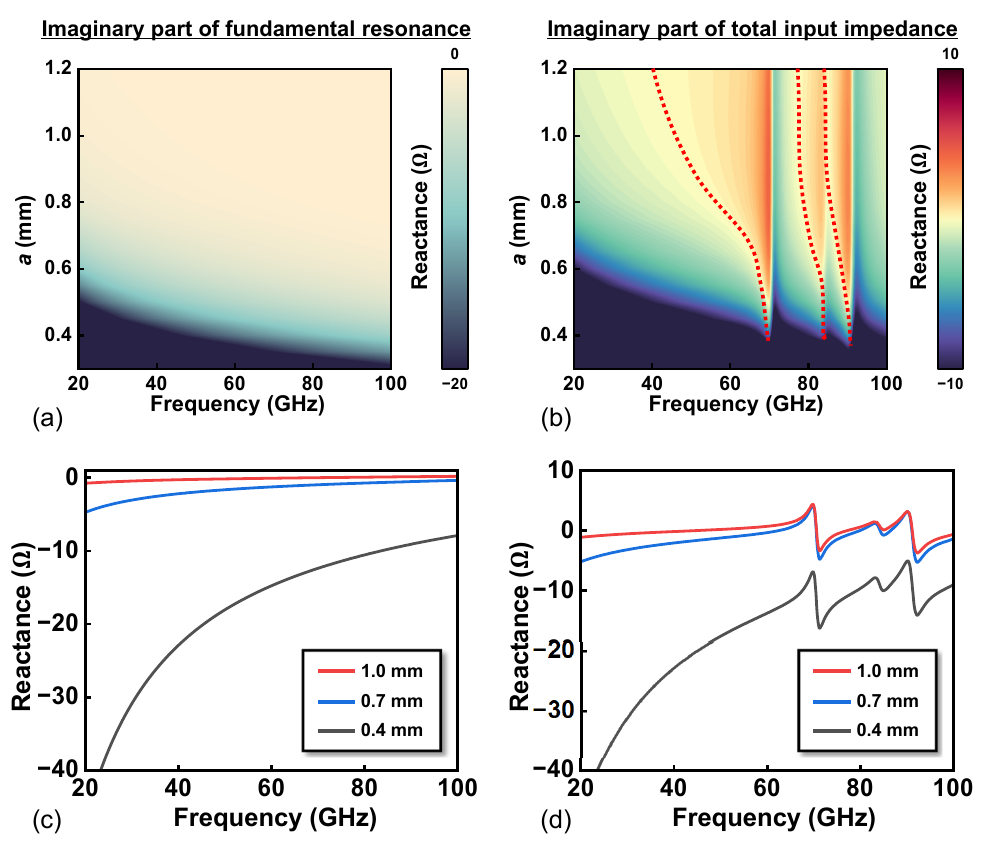}
\caption{Resistances of (a) the fundamental resonance and (b) total input impedance ($Y_\mathrm{in}$) regarding to the square pattern size ($a$) and the overall frequencies. Reactance components of (c) the fundamental resonance and (d) total input impedance ($Y_\mathrm{in}$) for certain values of the parameter $a$.}
\label{fig:Impedance_square}
\end{figure}

As shown in Fig. \ref{fig:Impedance_harmonics}(a) and (c), the variation of the square pattern size is not related to the singularity positions of the resistances for the $h$-th harmonics at all, since the parameters are only determined by the periodicity.
In terms of the reactance part, Fig. \ref{fig:Impedance_harmonics}(b) and (d) show the zero-crossing points are invariant by changes of $a$.
Meanwhile, the reactance of the fundamental resonance caused by the metasurface is widely spread within the overall frequencies for $a \le 0.7$ mm, as shown in Fig. \ref{fig:Impedance_square}(a) and (c).
Fig \ref{fig:Impedance_square}(b) and (d), the total input impedance less than the certain size of $a$ is decided by considering the reactance variation of the fundamental resonance, where the tendency is similar to the variations of the peak points in Fig. \ref{fig:Dielectric_case_square_size_HFSS_ECN}. 
The flat reactance for $a > 0.7$ mm represents that the fundamental resonance loses its dominance in the periodic structure, due to the increasing coupling of the higher order of the space harmonics.
In terms of the conditional size of $a$ and $P$, the fundamental and higher-order resonances of the metasurface can be controlled by independent parameters corresponding to each resonance.

\subsection{Effect of Material Loss for Trapped Surface Waves}

\begin{figure}
\centering
\includegraphics[width=1\linewidth]{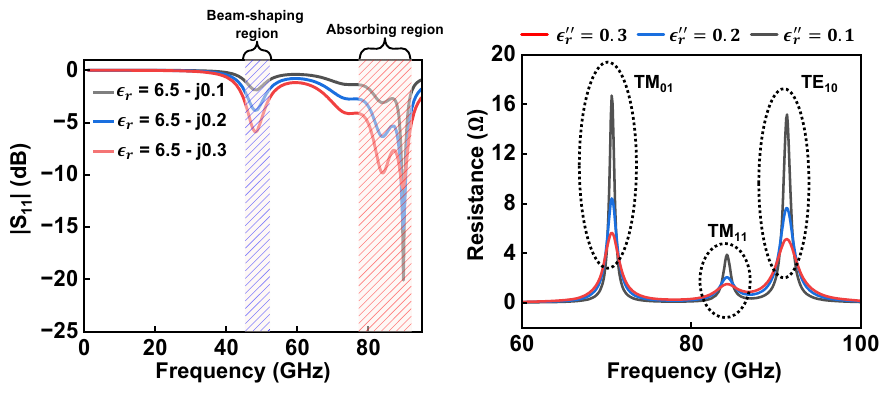}
\caption{(a) Reflection coefficient of the metasurface based on a dielectric material varies as increasing $\mathrm{Im} \{ \epsilon_r \}$ of the permittivity ($\epsilon_r$). (b) The resistance component of ($Z_\mathrm{eq}$) corresponding to the critical harmonics, is analyzed for cases where the imaginary part ($\epsilon_r''$) of the permittivity takes values of 0.1, 0.2, and 0.3.}
\label{fig:Loss_variation}
\end{figure}

Material losses, such as dielectric loss, are major parameters in including the overall attenuation over the trapped surface waves distributed in the periodic structure, as illustrated in Fig. \ref{fig:Dielectric_E_z_distribution}(a). 
Increasing a dielectric loss is effective to increase the quality factor of the reflection response for the $h$-th harmonics and strengthen the absorption property, as illustrated in Fig. \ref{fig:Loss_variation}(a).
The relationship can be specified by the periodic waveguide shorted at the back ($z = -d$), considering no coupling between the higher-order modes inside the cavity.
For the $z$-direction propagating component inside the periodic waveguide, the transverse electric field of $\bar{\mathbf{e}}_h (x, y)$ excited by the metasurface is given by
\begin{align}
\bar{\mathbf{E}}_h (x, y, z) = N_h \bar{\mathbf{e}}_h (x, y) \left(A^+ e^{-j \beta_{z,h}^{(a)}z}+ A^- e^{j \beta_{z,h}^{(a)}z}\right)
\end{align}
where the coefficient of $A^+$ and $A^-$ are arbitrary amplitudes of the forward and backward traveling waves. 
Using the condition that $\bar{\mathbf{E}}_h (x, y, 0) = N_h \bar{\mathbf{e}}_h (x, y)$ implies that $A^+ = 1 - A^-$ on the metasurface.
The Floquet harmonics become evanescent and exponentially decay in the free-space.
Then, the condition $\bar{\mathbf{E}}_h = 0$ at $z = -d$ leads to the coefficient 
\begin{align}
A^- &= \frac{e^{j \beta_{z,h}^{(a)}}}{e^{j \beta_{z,h}^{(a)} d} - e^{-j \beta_{z,h}^{(a)} d}}.
\end{align}
Considering the attenuation inside the dielectric/magnetic substrate of each periodic boundary, the transverse fields for the Floquet modes can be written as
\begin{align}
\bar{\mathrm{E}}_h (x, y, z) &= N_h \bar{\mathbf{e}}_h (x, y) \frac{e^{j \dot{\beta}_{z,h}^{(a)} (z + d)} - e^{-j \dot{\beta}_{z,h}^{(a)} (z + d)}}{e^{j \dot{\beta}_{z,h}^{(a)} d} - e^{j \dot{\beta}_{z,h}^{(a)} d}} \nonumber \\
&= N_h \bar{\mathbf{e}}_h (x, y) \underbrace{e^{-j \dot{\beta}_{z,h}^{(a)} (z + d)}}_\textrm{traveling wave} \underbrace{{\frac{1 - e^{2j\dot{\beta}_{z,h}^{(a)}(z + d)}}{2j \sin{\dot{\beta}_{z,h}^{(a)}d}}}}_\textrm{standing wave}, \\
\dot{\beta}_{z,h}^{(a)} &= \beta_{z,h}^{(a)} \left[ 1 - j\frac{\tan{\delta_\epsilon + \tan{\delta_\mu}}}{2 \{\beta_{z,h}^{(a)}\}^2} \{ \beta_{z, 0}^{(a)} \}^2 \right]
\label{eq:approx_wavenumber}
\end{align}
where the complex wavenumber considering the material loss is denoted as $\dot{\beta}_{z,h}^{(a)}$. The attenuation part within the traveling wave leads to decay in the magnitude of the propagating normal waves of all the harmonics, which is reflected in decreasing the resistance of each harmonic as shown in Fig. \ref{fig:Loss_variation}(b).
For a higher loss, the reflection response with a rapidly change for the higher frequency band can be designed in terms of the absorption, as suppressing the impedance of each harmonic contributes to extending the impedance matching region. 
It is notable that the variation of the loss properties has an effect on improving the quality factor of the periodic structure resonance, as well as that of the fundamental resonance.

\section{Novel Absorbing Property Based on Frequency Selective Material}

\subsection{Novel Absorbing Mechanism Combining Trapped Surface Waves and Magnetic Substrate}

\begin{figure}
\centering
\includegraphics[width=1\linewidth]{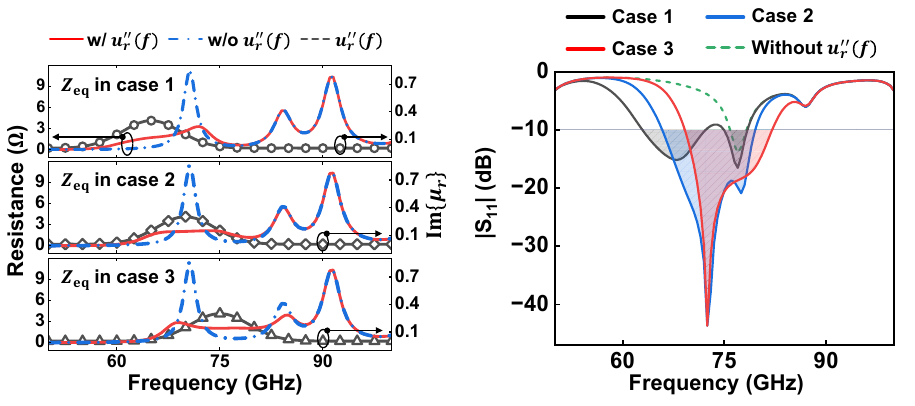}
\caption{(a) Resistance part of $Z_\mathrm{eq}$ considering an arbitrary function (\ref{eq:arbitrary_function}) with respect to $\mathrm{Im}\{\mu_r\}$: Case 1 where $f_c = 65$ GHz, case 2 where $f_c = 70$ GHz, and case 3 where $f_c = 75$ GHz. (b) Reflection cofficient of the metasurface with a virtual material comparing to the dielectric material, where an arbitrary function (\ref{eq:arbitrary_function}) of $\mathrm{Im}\{\mu_r\}$ is applied.}
\label{fig:Loss_function}
\end{figure}

When using conventional dielectric substrates, the absorption band appears in a narrow region characterized by rapid phase changes corresponding to a higher quality factor.
Furthermore, an increase in dielectric loss can impede the effective design of both the fundamental resonance and the periodic structure resonances. 
The growth of reflection loss adversely makes the reflection surface design in the lower operation challenging, as it severely degrades the propagation within the designated region.
Besides, the dielectric substrates faces inherent physical limitations in achieving a higher dielectric loss itself.
To overcome the physical limitations associated with dielectric substrates, the introduction of new materials, such as those varying in magnetic loss properties, is necessary \cite{gorbachev2022high, liu2016zr, dong2014microwave, liu2017excellent}.
As shown in Fig. \ref{fig:Loss_function}(a), a virtual material is assumed in introducing a variation of certain magnetic losses such as \cite{iijima2000millimeter}.
The imaginary part ($\mu_r''$) of the permeability depending on the frequency variation can be given as
\begin{align}
\mu_r''& = p \times e^{-k (f - f_c)^2}, 
\label{eq:arbitrary_function}
\\
k &= \frac{\ln 2}{(f_\mathrm{half} - f_c)^2}
\end{align}
where $p$ is denoted as the peak amplitude of $\mu_r''$, and the order ($k$) is determined by the half-magnitude bandwidth, from $f_c - f_\mathrm{half}$ to $f_c + f_\mathrm{half}$, adjacent to the center frequency ($f_c$).
As shown in Fig. \ref{fig:Loss_function}(a), the properties of the material can be reflected as a decrease of the input impedance ($Z_\mathrm{eq}$) within a specific region.
If a material exhibits magnetic loss within the frequency band for which the trapped surface waves propagates, the previously undesignated region can be transformed into an absorbing region, as shown in Fig. \ref{fig:Loss_function}(b).
The roll-off property or absorption bandwidth of the absorbing region can be designed by manipulating the magnetic loss function and the reflection response of the $h$-th harmonics.
Case 3 is possible to achieve the sharper roll-off in front of the start of the absorption band comparing to the other cases.
Meanwhile, the absorption band of the case 1 has a dual-band operation.

\begin{figure}
\centering
\includegraphics[width=1\linewidth]{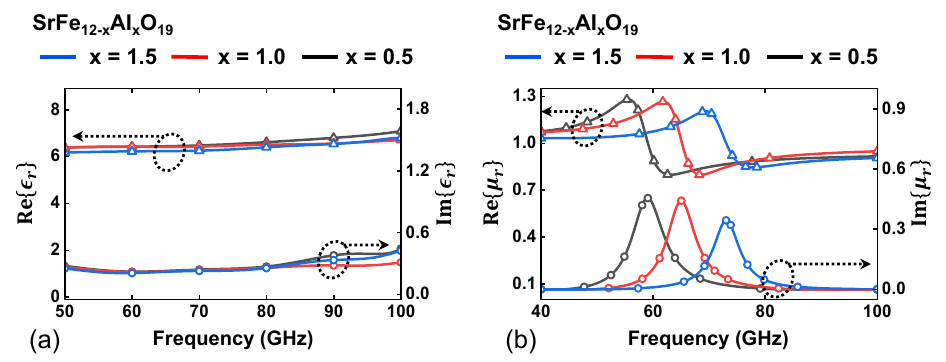}
\caption{(a) Permittivity and (b) permeability of the M-type ferrrite depending on the doping concentration of Al ion.}
\label{fig:Ferrite_properties}
\end{figure}
\begin{figure}
\centering
\includegraphics[width=1\linewidth]{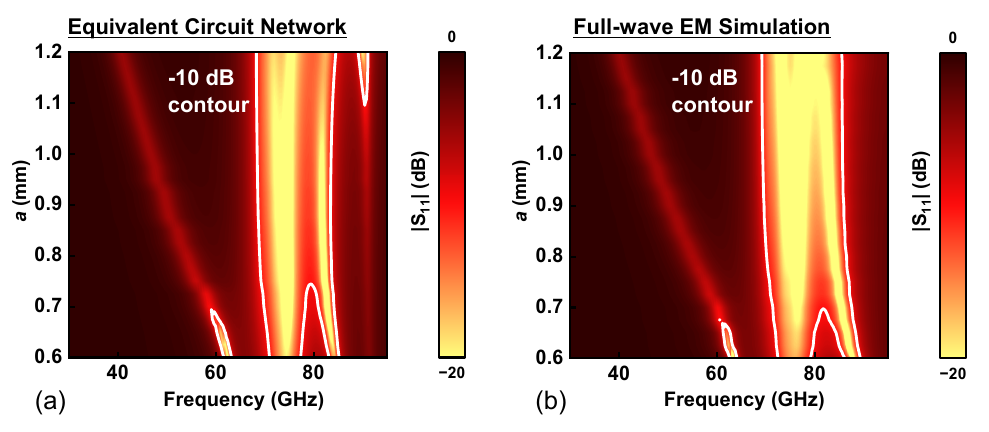}
\caption{Reflection cofficient of the metasurface with periodic structure based on the M-type ferrite as a function of the square pattern ($a$) calculated by (a) 3D FEM simulation and (b) ECN model approach.}
\label{fig:Ferrite_case_patch_HFSS_ECN}
\end{figure}
\begin{figure}
\centering
\includegraphics[width=1\linewidth]{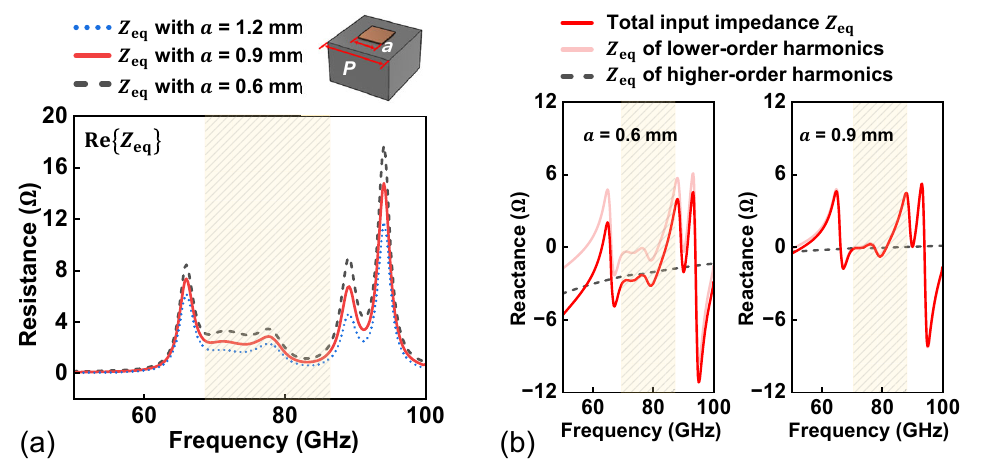}
\caption{ (a) Resistance part of $Z_\mathrm{eq}$ considering the different sizes of the patch pattern. (b) Reactance part of $Z_\mathrm{eq}$ considering the impact of the fundamental resonance and the periodic structure resonances for the $h$-th harmonics. It is clear that the pattern size needs to be set as 0.9 mm to achieve the continuous impedance matching within a desired region.}
\label{fig:Impedance_ferrite}
\end{figure}

To realize the virtual material and the loss function, M-type ferrites with a loss function simliar to the virtual material need to be designed and fabricated.
As shown in Fig. \ref{fig:Ferrite_properties}(a) and (b), the ferromagnetic resonance (FMR) of the ferrite substrate can be adjusted by controlling the doping concentration of Al ion \cite{lee2023absorption}, while the permittivity of that remains invariant.
Utilizing this tunable property, the FMR frequency of the M-type ferrite material is closely aligned with the transverse resonances of $f_\mathrm{tr1}$ and $f_\mathrm{tr2}$, which achieves a continuous absorption band such as Case 3 in Fig. \ref{fig:Loss_function}(b).
As shown in Fig. \ref{fig:Ferrite_case_patch_HFSS_ECN}(a) and (b), the integration of the ferrite substrate and the metasurface with periodic structure enhances the absorption bandwidth compared to a single slab of M-type ferrite without a metasurface.
Interestingly, the metasurface with a ferrite substrate maintains a stable operating bandwidth of its absorption, although the patch size $a$ varies between 0.7 and 1.2 mm.
This is because the square width change more than a certain size is capable of exciting the normal propagating waves of the Floquet harmonics with a coupling ratio of the comparable intensity for the fundamental resonance, as shown in Fig. \ref{fig:Impedance_square}(a) and (c).
In terms of the equivalent resistance in Fig. \ref{fig:Impedance_ferrite}(a), the transverse resonances caused by the trapped surface waves are located at constant frequenies for varying the pattern size ($a$).
Utilizing the M-type ferrite under the metasurface allows us to specify the continuous bandwidth for the absorption property.

\begin{figure}
\centering
\includegraphics[width=1\linewidth]{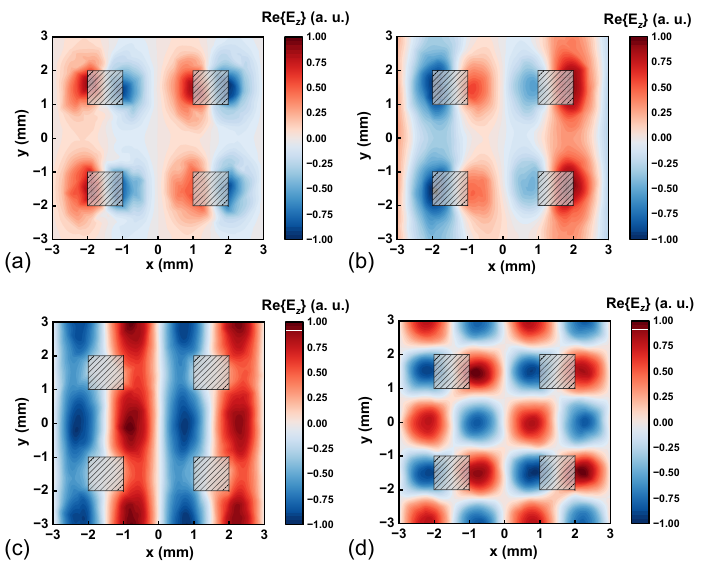}
\caption{Simulated $E_z$-field distributions at a few selected frequencies of (a) 48, (b) 60, (c) 70, and (d) 85 GHz in case of the M-type ferrite substrate. At frequencies of 48 and 60 GHz, the $E_z$-field distribution is concentrated around the periphery of each meta-unit. The $E_z$-field is spatially distributed within the surface lattice at 70 and 85 GHz, which is derived from the influence of the trapped surface wave of the $h$-th harmonic.}
\label{fig:Ferrite_E_z_distribution}
\end{figure}

The $E_z$-field distribution should be addressed to specify the operation of the fundamental resonance and periodic structure resonances.
As shown in Fig. \ref{fig:Ferrite_E_z_distribution}(a) and (b), the $E_z$-field distribution of the fundamental resonance for the metasurface with periodic structure is similar with that observed within the dielectric substrate.
The field distribution for a change of the metasurface substrate is invariant, located around the edge of each meta-unit.
The M-type ferrite based metasurface in the lower frequency can be designed into a reflective surface for tiliting the incident wave.
The $E_z$-field distribution of Fig. \ref{fig:Ferrite_E_z_distribution}(c) and (d) shows that the periodic structure resonances in the higher frequencies become intensive based on the M-type ferrite, comparing to the dielectric substrate.
In case of the M-type ferrite, the surface current is hard to be induced on the metasurface within the higher frequencies due to the mode matching between the fundamental resonance and the periodic structure of the Floquet harmonics.
The $E_z$-field distributions of the Floquet harmonics propagate along the metasurface inside the periodic wall without the fundamental resonance of the metasurface.

\subsection{Controlling Ferromagnetic Resonance Frequency of M-type Ferrite}

\begin{figure}
\centering
\includegraphics[width=0.5\linewidth]{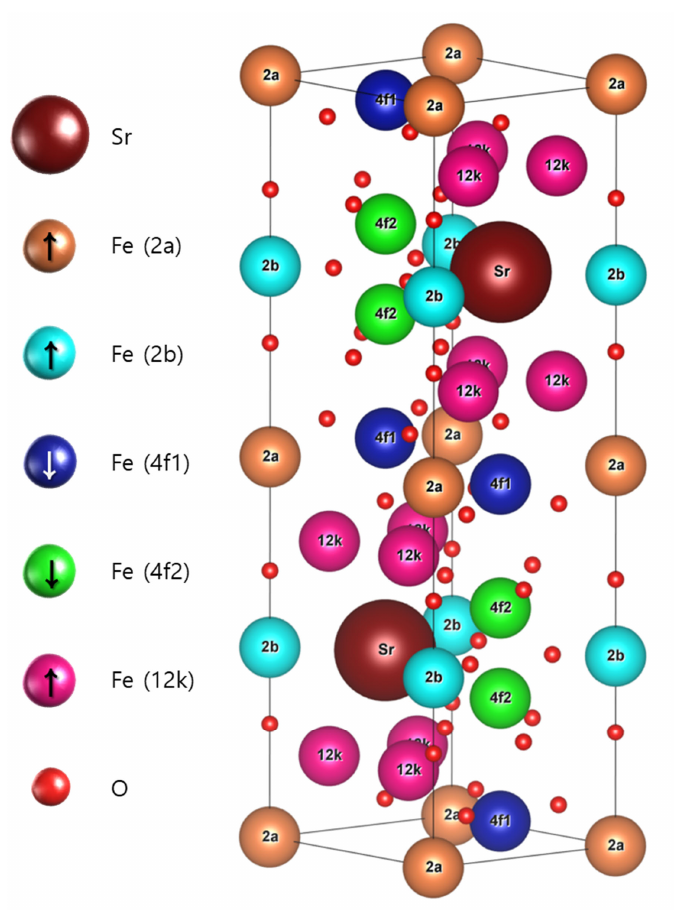}
\caption{Crystallographic structure of M-type strontium ferrite, containing five different Fe sites: (1) octahedral 2a site with up-spin, (2) trigonal bipyramidal 2b site with up-spin, (3) tetrahedral 4f1 site with down-spin, (4) octahedral 4f2 site with down-spin and (5) octahedral 12k site with up-spin. Due to the antiparallel arrangement of spins, the $M_s$ and $H_a$ of M-type strontium ferrite are affected in different ways depending on which site the localized Fe$^{3+}$ ions occupy.}
\label{fig:Fe_structure}
\end{figure}

The new material needs to be introduced in order to overcome the physical limitation when using the dielectric substrate. 
If a magnetic material has a magnetic loss in the frequency band that traps higher-order modes, the undesignated region can be changed to the absorbing region, as shown in Fig. \ref{fig:Application_description}. 
This desired high magnetic loss in a specific frequency band can be induced from ferromagnetic resonance (FMR) of the ferromagnetic materials. 
While the magnetic materials induce a significant loss of the electromagnetic wave at the FMR frequency via coupling between the wave and the magnetization of the ferromagnetic material, most magnetic materials including Fe, Co, and spinel ferrites have a low FMR frequency below several GHz \cite{snoek1948dispersion}. 
However, M-type ferrites are known to exhibit FMR at over 45 GHz due to their high magnetic anisotropy ($H_a$) compared to other magnetic materials, as the FMR frequency is proportional to $H_a$ of the magnetic material \cite{ratnam1970nature}.

The authors' group has been interested in controlling the FMR frequency of M-type ferrites by substituting Fe ions into various transition metal ions and inducing increased/decreased $H_a$. 
According to previous research by the authors, substitution of Co-Ti in M-type strontium ferrites (SrFe$_{12}$O$_{19}$) decreases the frequency of FMR while Al increases the FMR frequency \cite{lee2023absorption}. 
There are five different lattice sites for Fe ions in M-type ferrites: 2a, 2b, 12k, 4f1, and 4f2 (Figure \ref{fig:Fe_structure}). During ion substitution, the Co-Ti ions prefer to occupy the 4f2 and 2b sites, while the Al ions prefer to occupy 12k and 2a sites \cite{zhou1991site, przybylski2015mossbauer}. 
As each site has different contribution to the ferrite’s magnetic properties including saturation magnetization ($M_s$), coercive field ($H_c$), and magnetocrystalline anisotropy constant ($K_1$), we can observe $H_a$ and FMR frequency changes \cite{xu1983magnetic}. 
For example, the substitution of nonmagnetic Al to magnetic Fe ions results in a decrease $M_s$ and a rise $K_1$, leading to an increase $H_a$ and a blueshift of the FMR frequency.

In this study, to absorb the 77 GHz band, SrFe$_{10.5}$Al$_{1.5}$O$_{19}$ ferrites were synthesized by substituting 1.5 Fe ions of SrFe$_{12}$O$_{19}$ into 1.5 Al ions during the M-type strontium ferrite synthesis process. 
The detailed synthesis procedure is described in the previous work of the authors \cite{lee2023absorption}. 
A ferrite composite solution was prepared by mixing 70 wt\% of SrFe10.5Al1.5O19 with 30 wt\% of TPU solution using a planetary mixer (ARE-310, Thinky). 
The composite solution was cast into 100 $\mu$ m thick layers by bar coating. 
A magnetic substrate was obtained by stacking these composite layers at a desired thickness and pressing at 120 ℃ with 10 MPa for 20 minutes.

\begin{figure*}
\centering
\includegraphics[width=0.8\linewidth]{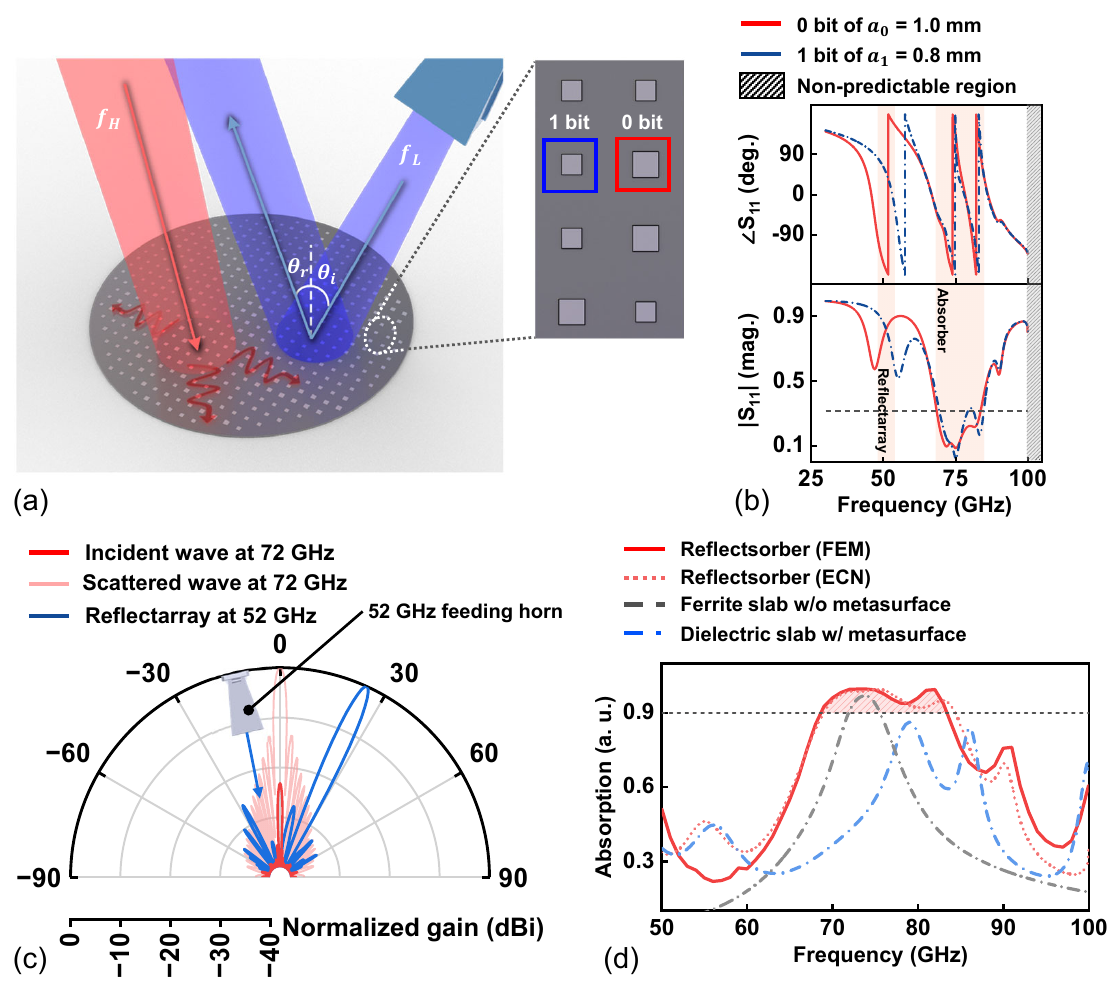}
\caption{(a) Multi-functional metasurface with the reflectarray and the absorbing property. (b) Reflection coefficient of 1-bit unit-cells with the same absorption region, representing that the phase variation of the dominant mode is decided by $a$. (c) Scattered beam-patterns of the reflectarray and the absorber functions corresponding to the lower frequency band and the higher frequency, respectively. The incident wave of the reflectarray and the absorber are designed as a feeding horn of 53 GHz and a incident plane wave, respectively. (d) Absorption performance of the metasurface under the normal incidence. The non-designated region can be changed into the absorption region, which is only achieved by combining the metasurface with the ferrite material.}
\label{fig:Reflectsorber}
\end{figure*}

\section{Reflectsorber Based on M-type Ferrite}

\subsection{Design}

The combination of the M-type ferrite slab with the metasurface offers distinct frequency-division properties, preserving the designated operation of the metasurface with periodic structure in the lower frequency bands.
When the field distributions are compared in Figs. \ref{fig:Dielectric_E_z_distribution} and \ref{fig:Ferrite_E_z_distribution}, the ferrite substrate primarily is related to higher frequencies and does not affect the metasurface operation in the lower bands, which stems from its material characteristics. 
The M-type ferrite substrate behaves similarly to a dielectric material without variations in relative permeability, that is, $\mu_r = 1$, as the permeability property of M-type ferrite is not able to be sustained below its FMR frequency.
The variation of the metasurface pattern, satisfying a proper ratio of $a$ to $P$, determines a phase variation of the fundamental resonance in the lower frequency region.
The beam-modulating designs, such as beam-shifting, anomalous reflection, and beam-scattering, can be applied to the lower-band operation of the metasurface.
In addition, a fixed periodicity of the metasurface based on the M-type ferrite provides a constant absorption band even on the variation of the pattern size ($a$).  

A simple device, referred to synthetically as a \textit{reflectsorber}--a term that encapsulates its dual functionality as both a reflectarray and an absorber--is proposed and shown in Fig. \ref{fig:Reflectsorber}(a).
The sheet impedances ($Y_\mathrm{es}$ and $Z_\mathrm{ms}$) of the metasurface are given by the reflection coefficient ($\Gamma$), since the transmittance is effectively zero due to the ground plane at the bottom of the metasurface \cite{pfeiffer2013metamaterial}.
Also, This configuration ensures that the reflection coefficients in both the $x$- and $y$-directions ($\Gamma = \Gamma_{xx}$ = $\Gamma_{yy}$) are equal, due to the symmetric shape of the elements.
The reflectarray metasurface can be designed by tailoring the reflection phase of each meta-unit based on conventional array theory~\cite{nayeri2013radiation}.
To realize beam steering with a 1-bit coding scheme, the elements must provide two distinct reflection phases—typically $0^\circ$ and $180^\circ$.
In Fig.~\ref{fig:Reflectsorber}(b), the 1-bit meta-atoms designed with lateral dimensions of $a_0 = 1.0$ mm and $a_1 = 0.8$ mm yield a phase difference of $180^\circ$.
The radiation pattern in Fig. \ref{fig:Reflectsorber}(c) illustrates that the metasurface works as the reflectrarray with the horn antenna operating in the U-band.

\begin{figure*}
\centering
\includegraphics[width=1\linewidth]{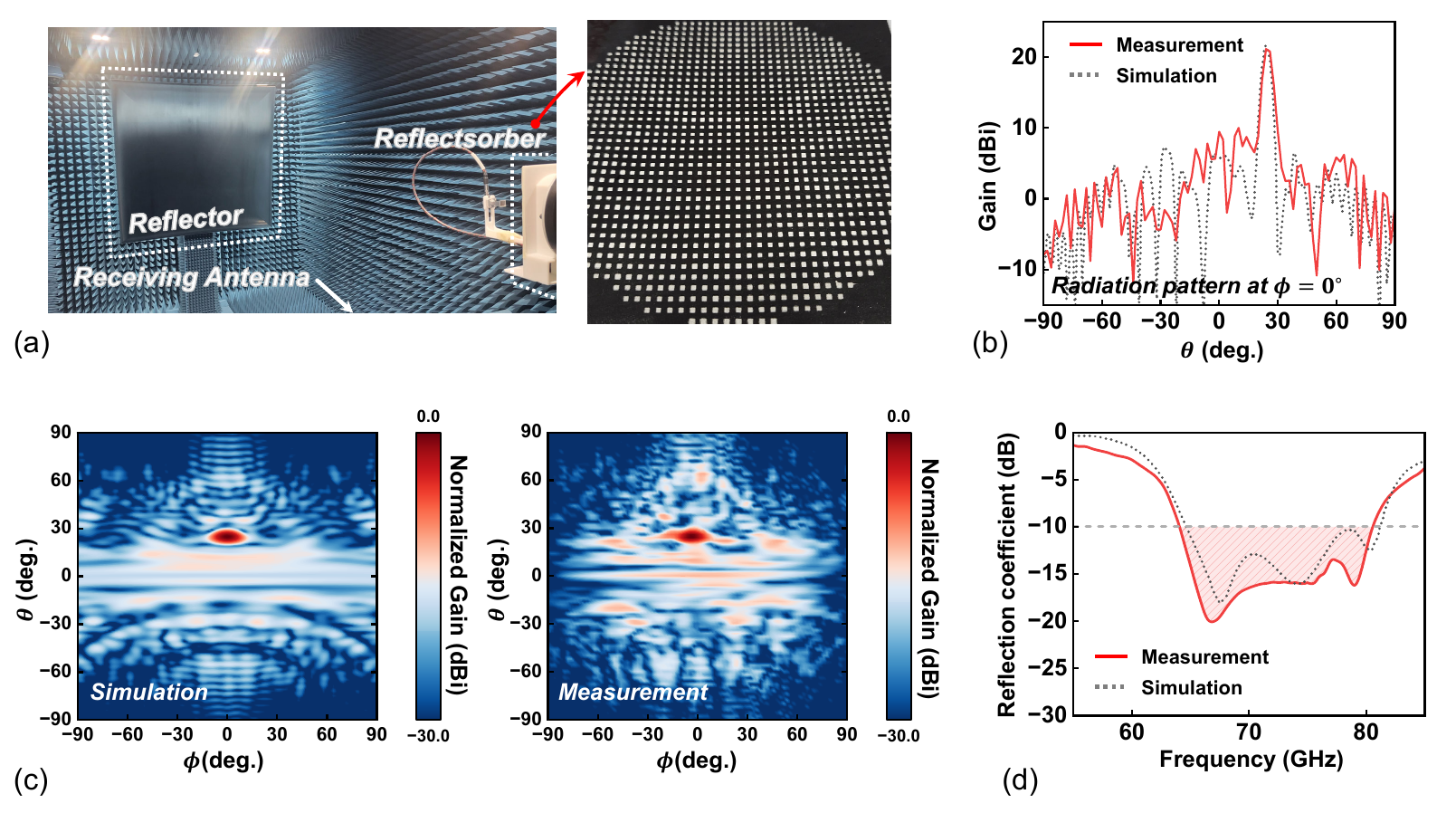}
\caption{(a) Far-field experimental setup and fabricated reflectsorber sample to measure the radiation performance of the reflectsorber. (b) Simulated and measured realized gain of the reflectsorber at $\phi = 0^\circ$ and 39 GHz. (c) Contours of simulated and measured realized gain of the reflectsorber at 39 GHz. (d) Absorption performance comparing the simulation with the measurement under the normal incidence.}
\label{fig:Reflectsorber_measurement_re}
\end{figure*}

The unit cell periodicity should satisfy the condition of $P < \lambda_0 / 2$ within the lower frequencies, which guarantees no grating lobe and continuous phase distribution.
The periodicity of 3 mm is enough to ensure the radiation performance of the metasurface.
In addition, a constant absorption band can be achieved by combining the metasurface with the M-type ferrite in the case of $P = 3$ mm in a higher frequency operation.
As shown in Fig. \ref{fig:Reflectsorber}(c), the absorbing property in the higher frequency is validated by the degraded intensity of the scattered radiation for the incident wave.
The absorbing property is designated in the higher operating band despite considering the reflectarray design within the metasurface, as depicted in Fig. \ref{fig:Reflectsorber}(d).
The FMR of the M-type ferrite is adjustable in V-band by modifying the Al ion concentration, as demonstrated in Supplementary Fig. S6. 
Once the operation band of the absorption capability is confirmed, the center frequency of the reflective region is defined by the resonance frequency of the 1-bit elements. 
For the reflectsorber with a periodicity of 3 mm, the operation frequency of the reflective regime ranges from 38 to 52 GHz by adjusting the patch size from 0.8 to 1.2 mm, as shown in Fig. \ref{fig:Ferrite_case_patch_HFSS_ECN}.

\subsection{Measurement Results}

To validate the theoretical approach of the reflectsorber, the metasurface was designed with a reflectarry operation of 38 GHz and the absorption operation of 78 GHz, which is based on the 1-bit unit cell of $a_0 = 1.1$ mm and $a_1 = 1.3$ mm in $P = 3.2$ mm.
The screen printing method was used directly to fabricate the metasurface printed on the M-type ferrite.
Fig. \ref{fig:Reflectsorber_measurement_re}(a) shows the fabricated reflectsorber based on the M-type ferrite slab with the FMR frequency of 70 GHz.
The designed absorber consists of a circular pattern with a radius of 20 unit cells, each measuring $0.41\times0.41 \lambda_0^{2}$.

In the first measurement, the radiation performance of the reflectsorber was tested in the anechoic chamber, as shown Fig. \ref{fig:Reflectsorber_measurement_re}(a).
The main peak of reflection toward the receiving antenna occurs at $\theta = 25^\circ$.
This is an expected result, meaning that the designed metasurface works as a reflectarray.
Material losses from screen printing for patterning the metasurface contribute to a 0.5 dB difference predicted between the peak gain.
The measured pattern of the $\phi = 0^\circ$ cut plane at 39 GHz agrees well with the simulation, as depicted in Fig. \ref{fig:Reflectsorber_measurement_re}(b).
Also, the contour pattern in Fig. \ref{fig:Reflectsorber_measurement_re}(c) represents the radiation performance similar to the expectation was exactly realized in utilizing the M-type ferrite and the metasurface with periodic structure.
For applications where a circular polarization is required, the metasurface is capable to radiate the circularly polarization by using a circularly polarized feed antenna.

\begin{figure*}
\centering
\includegraphics[width=1\linewidth]{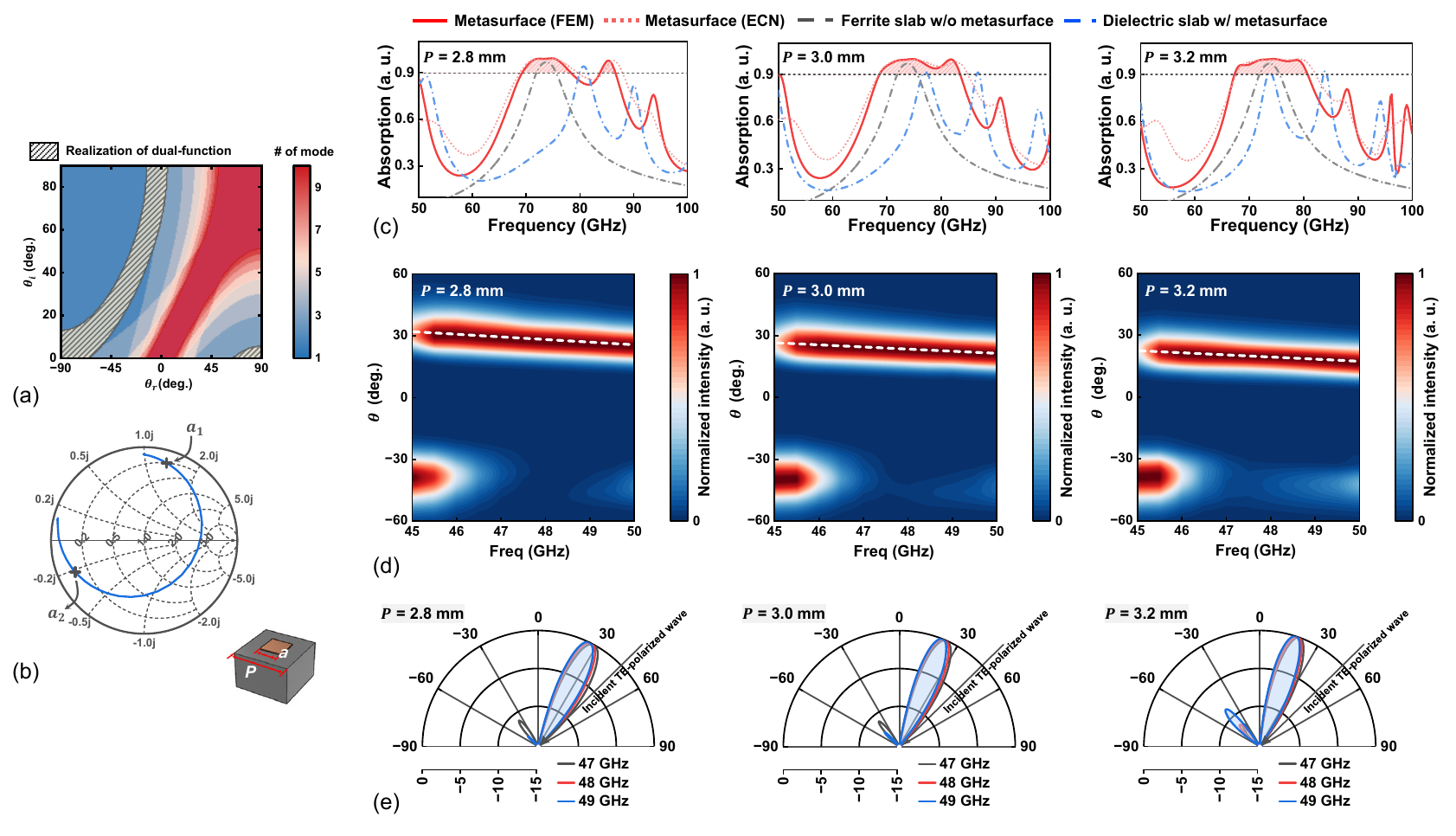}
\caption{(a) The number $(M)$ of outgoing diffraction modes and the realization region in supporting the dual-operation based on the ferrite substrate with the FMR of 74 GHz. (b) The reflection coefficient for varying $a$ is represented on the Smith chart, and the parameters $(a_1, a_2)$ are selected for the metasurface design. (c) Absortpion performance as the unit-cell size of metasurface varies under the normal incidence. The non-designated region in the higher-frequency band can be changed into an absorption region, which is only achieved by combining the metasurface with the ferrite material. (d) Simulated scattering of a finite metasurface as the unit-cell size ($P$) varies. The desired reflected angle can be tuned by the unit-cell size. (e) The corresponding 1-D normalized intensity on th e frequencies of 47, 48, and 49 GHz.}
\label{fig:Anomalous_reflection}
\end{figure*}

In the second experiment, the absorption property can be validated by the reflection ratio for the metasurface under the incident wave compared to that of the reference metal plate.  
To measure the scattering parameters of the absorbers, a vector network analyzer (Keysight N5291A) was used with a focusing lens antenna (FS-110, EM Labs) for the free space measurement method in the range of 18 $\sim$ 110 GHz.
A time-gating technique was implemented to eliminate background noise and other potential effects, retaining only the signal reflected from the metasurface.
Fig. \ref{fig:Reflectsorber_measurement_re}(d) shows the measurement of absorption performance under normal incidence at V- and W-bands.
As expected in both the theoretical and the simulated results, the metasurface can be spanning 63 $\sim$ 82 GHz with a fractional bandwidth of 26.2\%, encompassing a commercial radar frequency of 77 GHz band.
The proposed absorption mechanism can be verified by the measurement, which matches well with the simulation.
The metasurface has the relative thickness of 0.083$\lambda_0$, where $\lambda_0$ is the free-space wavelength at the low frequency in the operating band.
By adhering to Rozanov's limitation, the absorption bandwidth ratio to the structure's thickness is constrained, which means that thinner absorbers tend to have narrower absorption bandwidth \cite{rozanov2000ultimate}. 
For an actual application, the reflectsorber can be utilized to suppress the high level of harmonic radiation for feeding antenna, such as microstrip antenna.
Also, the absorption property of the reflectsorber is capable to work as a planar filter, which is effective to mitigate an external interference, without the additional design for introducing a filtering structure.

\begin{figure*}
\centering
\includegraphics[width=1\linewidth]{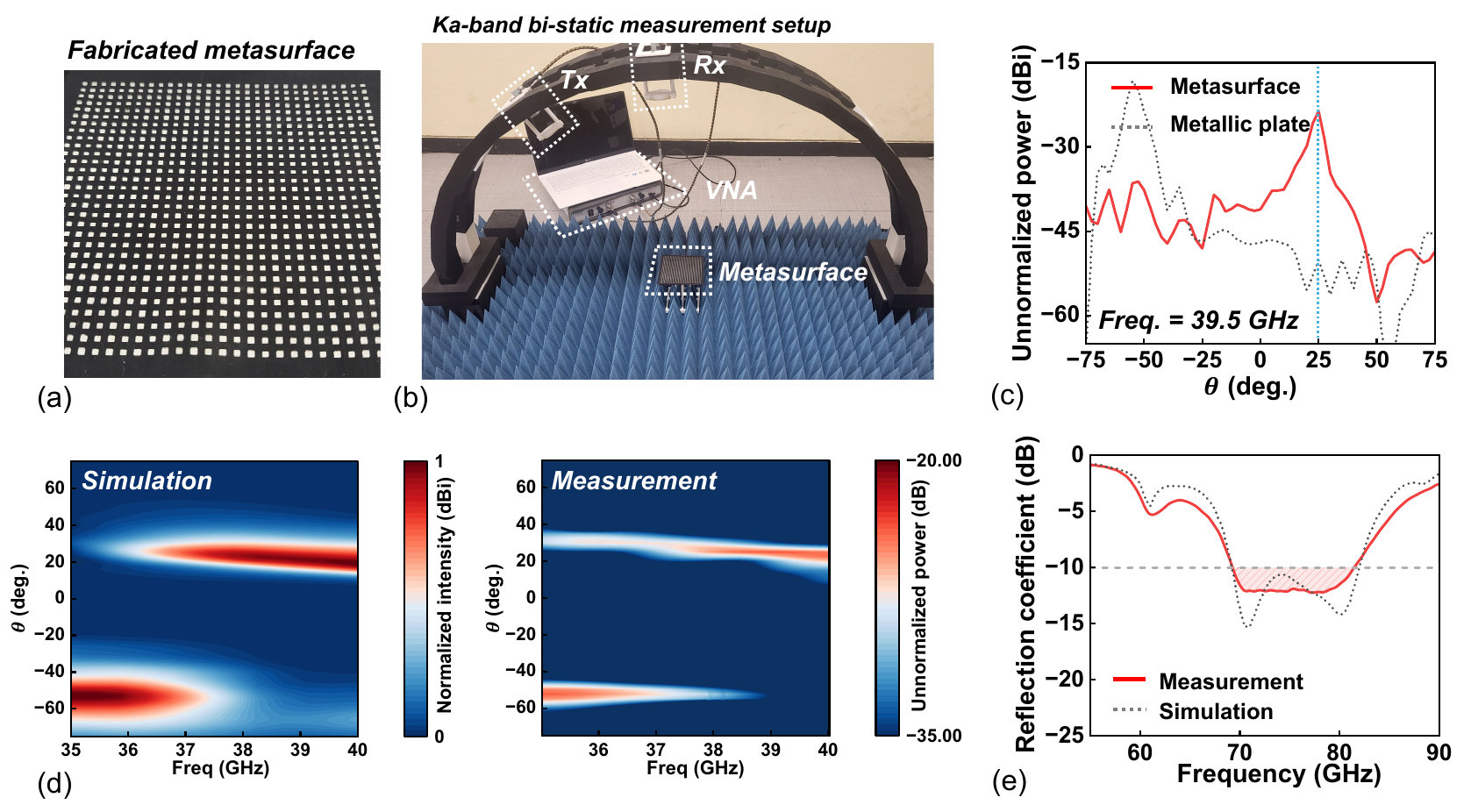}
\caption{(a) Fabricated metasurface sample. (b) Ka-band bi-static experimental setup in a semi-anechoic chamber to measure the radiation performance of the designed metasurface. (c) Measured signals when varying the orientation angle $\theta$ of the receiving antenna for the metasurface and the reference plate with its same size. (d) Contours of simulated and measured normalized intensity of the metasurface for reflected angle ($\theta$) and frequency band of 35--40 GHz. (e) Absorption performance comparing the simulation with the measurement under the normal incidence.}
\label{fig:Metasurface_measurement}
\end{figure*}

\section{Metasurface Combining Anomalous Reflective Behavior and Absorbing Property}

\subsection{Design}

An additioanl example of this configuration is to integrate the novel absortive property with anomalous reflectior analyzed in \cite{wong2018perfect}, demonstrating the anomalous reflective behavior of the metasurface realized by a 1-bit element for an incident plane wave.
The period of the metasurface that satisfies $k_g = k_0 (\sin{\theta_r} - \sin{\theta_i}) = 2 \pi / \lambda_g$ is crucial to support the anomalous reflection caused by the propagation of the diffraction modes.
As described in \cite{wong2018perfect}, the number of outgoing diffraction modes is determined by 
\begin{align}
M = 1 + \left \lfloor \frac{1 - \sin{\theta_i}}{|\sin{\theta_r} - \sin{\theta_i}|} \right \rfloor + \left \lfloor \frac{1 + \sin{\theta_i}}{|\sin{\theta_r} - \sin{\theta_i}|} \right \rfloor
\end{align}
where $\lfloor \, \cdot \, \rfloor$ indicates the floor (round down) operation.
In the case of the region shaded in blue, high-efficiency anomalous reflection is achievable with two elements per guided wavelength $\lambda_g$ ($M = 2$) \cite{wong2018perfect}, as shown in Fig. \ref{fig:Anomalous_reflection}(a).
For anomalous reflection, the set of incident angle ($\theta_i$) and reflected angle ($\theta_r$) is selected and designed within the shaded region.
The periodicity ($P = \lambda_g/2$) has to be meticulously considered to ensure effective operations of both the the designated region and the emergent absorption feature.
The chosen set of $\theta_i$ and $\theta_r$ should meet the following condition, which governs the absorbing properties of the metasurface:
\begin{align}
&\lambda^\mathrm{LO}_0/P_\mathrm{max} \le |\sin{\theta_r} - \sin{\theta_i}| \le \lambda^\mathrm{LO}_0/P_\mathrm{min}
\end{align}
where the free-space operating wavelength of the reflective mode in lower frequency is $\lambda^\mathrm{LO}_0$.
For the transverse resonances in Fig. \label{fig:Dielectric_case_unit_cell_size}(b), the maximum and minimum of the unit-cell size are given as
\begin{align}
f_\mathrm{tr1}(P_\mathrm{min}) \leq f_\mathrm{FMR} \leq f_\mathrm{tr2}(P_\mathrm{max})
\end{align}
where the continuous absorption band can be obtained by fixing the periodicity of the metasurface.
For example, when considering an M-type ferrite substrate with FMR frequency of 70 GHz, unit cell sizes are determined to be a minimum ($P_\mathrm{min}$) of 3.0 mm and a maximum ($P_\mathrm{max}$) of 3.6 mm.

The patterned area in Fig. \ref{fig:Anomalous_reflection}(a) shows the specified region desired to achieve the realization of the anomalous reflective property ($\lambda^\mathrm{LO}_0$) and the absorbing property ($\lambda^\mathrm{HI}_0$), when using a ferrite substrate with an FMR frequency of 75 GHz.
The metasurface integrated with a ferrite slab can be engineered to reflect a TE-polarized wave with 40$^\circ$ into an anomalous reflection with 23$^\circ$.
For a patch size where $a = b$, Fig. \ref{fig:Anomalous_reflection}(b) represents the reflection coefficient as $a$ is varied from 0.8 to 1.3 mm in $\lambda^\mathrm{LO}$.
There is a reflection loss caused by the dielectric loss of $\tan \delta_\epsilon = \epsilon_r'' / \epsilon_r'$, which has an impact on the overall frequency band.
The 1-bit elements with $a_1$ and $a_2$ are determined to be 0.85 mm and 1.03 mm, respectively, to achieve anomalous reflection.
The phase difference between the 1-bit elements needs to be 180$^\circ$ to reduce specular reflection.
Furthermore, with a fixed unit cell periodicity, the metasurface elements maintain similar frequencies for the absorbing operation in the higher-order mode, even when the patch size is varied.
As shown in Fig. \ref{fig:Anomalous_reflection}(c), the metasurface on a ferrite substrate with a FMR frequency of 73 GHz exhibits a broader absorption bandwidth under variations in $P$, with bandwidths of 11 GHz, 17 GHz, and 14 GHz at $P$ values of 2.8, 3.0 and 3.2 mm, respectively.
In constant, when only the ferrite slab is used, the bandwidth narrows significantly to only 2.8 GHz.
Specular reflection is suppressed across a frequency range from 47 to 49 GHz as $P$ varies from 2.8 to 3.2 mm.
For further reduction of specular reflection at $P = 3.2$ mm, the dimensions of the 1-bit elements need to be finely re-optimized from $a_1$ and $a_2$, considering deviations in a mutual coupling between 1-bit unit cells.
The reflected angle can be adjusted by tuning the guided diffraction modes, which change with the unit cell size, as illustrated in Fig. \ref{fig:Anomalous_reflection}(d) and (e) \cite{asadchy2017flat}.
The metasurface achieves an anomalous reflection with a power efficiency of 60\%, taking into account the inherent reflection loss of the substrate.

\subsection{Measurement results}

To validate the theoretical approach of the proposed metasurface, the modulated surface was designed with an anomalously reflective operation of 39 GHz and the absorption operation of 78 GHz, which is based on the 1-bit unit cell of $a_0 = 1.05$ mm and $a_1 = 1.2$ mm in $P = 3.2$ mm.
The screen printing method was used directly to fabricate the metasurface printed on the M-type ferrite.
Fig. \ref{fig:Metasurface_measurement}(a) shows the metasurface fabricated based on the M-type ferrite slab with a FMR frequency of 74 GHz.
The designed absorber consists of a square pattern with 30 unit cells, where each cell is measured as the electrical width $0.41\times0.41 \lambda_0^{2}$ corresponding to the lower-band operation.

In the first measurement, the reflection characterization of the metasurface was tested in the semi-anechoic chamber, as shown in Fig. \ref{fig:Metasurface_measurement}(b).
In the bi-static measurement setup, the 3D-printed lens horn antennas with 23 dBi at 39 GHz were used to collimate the energy for receiving and transmitting the electromagnetic wave. 
The metasurface was placed at a distance of 50 cm (about 65$\lambda$ at 39 GHz) from the antennas.
The distance is enough to satisfy the far-field condition, where the radiation from the antenna can be approximated as a plane wave.
The receiving antenna was rotating along the bridge in the range of $-75^\circ$ to $75^\circ$ with a setup of $5^\circ$.
The main peak of reflection toward the receiving antenna occurs at $\theta = 25^\circ$, as depicted in Fig. \ref{fig:Metasurface_measurement}(c).
This is an expected result, meaning that the designed metasurface works as an anomalous reflection.
The material losses of M-type ferrite and the surface roughness caused by the screen printing contribute to a difference of 4 dB compared to the metal plate.
In addition, the contour pattern in Fig. \ref{fig:Metasurface_measurement}(d) represents the reflection ability similar to the expectation was exactly realized, although the anomalous reflective metasurface was designed and fabricated on the M-type ferrite.

In the second experiment, the absorption property can be validated by the reflection ratio for the metasurface under the incident wave compared to that of the reference metal plate. 
To measure the scattering parameters of the absorbers, a vector network analyzer (Keysight N5291A) was used with a focusing lens antenna (FS-110, EM Labs) for the free space measurement method in the range of 18 $\sim$ 110 GHz.
A time-gating technique was implemented to eradicate background noise and other potential effects, retaining only the signal reflected from the metasurface.
Fig. \ref{fig:Metasurface_measurement}(e) shows the measurement of absorption performance under normal incidence at V- and W-bands.
As expected in both the theoretical and simulated results, the metasurface can be spanning 69 $\sim$ 83.5 GHz.
The proposed absorption mechanism can be verified by the measurement, which matches well with the simulation.

\section{Conclusion}

This paper has provided a comprehensive review and tutorial on multi-functional metasurfaces integrated with M-type ferrite materials for millimeter-wave (mmWave) absorption and beam control. 
As wireless communication moves toward beyond-5G systems and incorporates non-terrestrial networks (NTNs), adaptive, low-profile electromagnetic surfaces capable of managing interference and enabling dynamic beam reconfiguration have become essential. 
This work have highlighted the limitations of conventional metasurfaces, particularly their challenges in simultaneously achieving wideband absorption and beamforming due to inherent material and structural constraints.

The review emphasized recent advances in multi-functional metasurface designs, showcasing the integration of frequency-selective magnetic materials and periodic surface lattices to realize passive, compact, and reconfigurable electromagnetic devices.
Special attention was given to the novel metasurface based on M-type ferrite materials, demonstrating their significant role in enhancing absorption through ferromagnetic resonance, alongside surface-wave trapping mechanisms to achieve both narrowband and broadband performance.
A detailed case study of a ferrite-based hybrid "reflectsorber," combining reflectorarray and absorber functionalities, was presented to illustrate critical design principles, analytical methodologies, and practical applications relevant to advanced communications with NTNs such as satellite communications and unmanned aerial vehicles (UAVs). 

{\appendix[Equivalent Circuit Network Modeling]

\begin{figure}[t]
\centering
\includegraphics[width=1\linewidth]{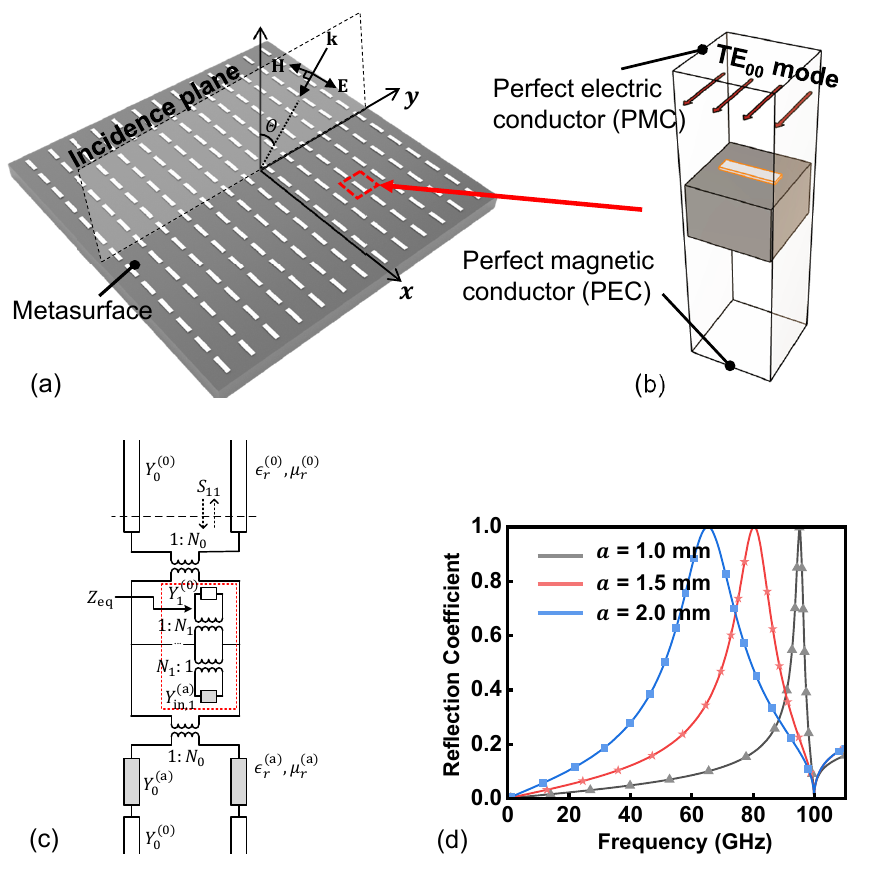}
\caption{(a) Scheme of the patch array of finite thickness under oblique incidence. (b) The patch array is equivalent to the waveguide discontinuity problem with a periodic boundary wall. (c) Equivalent circuit network modeling for computing the scattering parameter $S_{11}$. (d) Magnitude of the reflection coefficient for the structure inside the periodic wall under the normal incidence and the patch array with $b = 0.5$ mm and $P = 3$ mm, where the metasurface is on the free space.}
\label{fig:Equivalent_network_analysis}
\end{figure}

It is considered that the metasurface composed of zero-thickness metallic patches lies on either a dielectric or ferrite substrate. 
As shown in Fig. \ref{fig:Equivalent_network_analysis}(a), a TE-polarized plane wave is incident at an angle $\theta$ in the principal $yz$ plane of the patch array
The impinging field parallel to the periodic structure is in the $E$-plane, inducing the surface currents on the metasurface.
Reflection and transmission from the metasurface with periodic structure correspond to the interference of an infinite set of propagating and evanescent plane waves, i.e., Floquet harmonics.
As shown in Fig. \ref{fig:Equivalent_network_analysis}(b), the propagation of the incident and reflected plane waves can be modeled by the equivalent circuit network (ECN) by waveguide and the impedance matching concepts \cite{rodriguez2015analytical, mesa2018unlocking}.
When the plane wave arrives at the metasurface with periodic structure, the scattered harmonics can be excited along the surface lattice.
The wave numbers ($k_{x,m}$ and $k_{y,n}$) in the $x$- and $y$-axes associated with the Floquet harmonics of order $m$ and $n$ are denoted by
\begin{align}
&k_{x,m} = \frac{2m\pi}{P} \\
&k_{y,n} = k_0 \sin{\theta} + \frac{2n\pi}{P}
\end{align}
where $m$ and $n$ are integer numbers ($m, n = 0,\pm1,\pm2,\cdots$), and $k_0$ is the free-space wavenumber.
Inside a substrate, the normal wave-number of $h$-th harmonic with a mode pair of $mn$ are defined as follows:
\begin{flalign}
&k^{(a)}_{z,h} = 
	\begin{cases}
	\beta_{h}^{(a)} = \sqrt{\{ k^{(a)} \}^2 - |\mathbf{k}_{\mathrm{t},h}|^2}, & k^{(a)} \ge |\mathbf{k}_{\mathrm{t},h}| \\
	\ -j \alpha_{h}^{(a)} = -j\sqrt{|\mathbf{k}_{\mathrm{t},h}|^2 - \{ k^{(a)} \}^2}, &  k^{(a)} < |\mathbf{k}_{\mathrm{t},h}|
	\end{cases}
\label{eq:wavenumber}
\end{flalign}
where the type of substrate is indicated by the superscript ($a$). 
The superscripts ($\mathrm{d}$) and ($\mathrm{m}$) for ($a$) indicate the dielectric substrate with $\mu_r = 1$ and the magnetic substrate with $\mu_r \ne 1$, respectively.
If the superscript is ($0$), the substrate is considered to be the free space with $\epsilon_r = 1$ and $\mu_r = 1$.
For a lossless material, the harmonics are represented as purely real or purely imaginary wavenumbers, which indicates whether it is a propagating or evanescent wave.

As reported in \cite{rodriguez2015analytical, mesa2018unlocking}, this phenomenon is analyzed by the equivalent network modeling, which addresses the discontinuities between the substrate and metasurface as shown in Fig. \ref{fig:Equivalent_network_analysis}(a).
The metasurface can be coupled with various harmonics, while the incident harmonic ($m = 0$ and $n = 0$) is scattered on the dielectric/magnetic substrate and the metasurface as shown in Fig. \ref{fig:Equivalent_network_analysis}(b).
The Floquet expansion of the electric and magnetic field at the discontinuity plane ($z = 0$) is given by
\begin{align}
\mathbf{E}(x, y) & = (1 + R) \mathbf{e}_0(x, y) + \sum_{h \ne 0} V_h \mathbf{e}_h(x,y) \\
\mathbf{H}^{(0)}(x,y) & = Y_0^{(0)} (1 - R) [\hat{\mathbf{z}} \times \mathbf{e}_0 (x,y)] \nonumber \\
& \quad - \sum_{h \ne 0} Y_h^{(0)} V_h [\hat{\mathbf{z}} \times \mathbf{e}_h (x,y)] \label{eq:magnetic_field1} \\ 
\mathbf{H}^{(a)}(x,y) & = Y_0^{(a)} (1 + R) [\hat{\mathbf{z}} \times \mathbf{e}_0 (x,y)] \nonumber \\ 
& \quad - \sum_{h \ne 0} Y_h^{(a)} V_h [\hat{\mathbf{z}} \times \mathbf{e}_h (x,y)] \label{eq:magnetic_field2}
\end{align}
where $R$ is denoted as the reflection coefficient of the incident wave, and $V_h$ indicates the unknown coefficient of the electric field expansion.
In the expansion, both TM and TE harmonic modes are included, with the exception of the incident harmonic ($h = 0$).
The index $h$ represents a surrogate number of the various $(n, m)$ harmonics, each characterized by a wave vector $\mathrm{k}_{\mathrm{t}, h}$ and its normalized transverse field $\mathbf{e}_h(x, y)$. 
The wave vector is comprised of the Floquet expansion of the field.
As expressed in \cite{mesa2014circuit}, this wave vector and related variables are given by
\begin{align}
\mathbf{e}_h (x,y) &= \frac{e^{-j\mathbf{k}_{\mathrm{t},h} \cdot \mathbf{\rho}}}{\sqrt{P_xP_y}} \hat{\mathbf{e}}_h \quad [\mathbf{\rho} = x\hat{\mathbf{x}} + y\hat{\mathbf{y}}] \\
\mathbf{k}_{\mathrm{t},h} &= k_{x,m} \hat{\mathbf{x}} + k_{y,n} \hat{\mathbf{y}} \\
\hat{\mathbf{k}}_{\mathrm{t}, h} &= \frac{\mathbf{k}_{\mathrm{t},h}}{|\mathbf{k}_{\mathrm{t},h}|} 
= \frac{k_{x, m} \hat{\mathbf{x}} + k_{y, n} \hat{\mathbf{y}}}{\sqrt{k_{x, m}^2 + k_{y, n}^2}} \\
\mathbf{v}_{h} &= \begin{cases}
\hat{\mathbf{k}}_{\mathrm{t}, h}, & \text{TM harmonics}. \\
-\hat{\mathbf{z}} \times \hat{\mathbf{k}}_{\mathrm{t}, h}, & \text{TE harmonics}.
\end{cases}
\end{align}
The wave admittances, $Y_{h}^{(a)}$, looking into the magnetic substrate and air are given by
\begin{align}
Y_h^{(a)} = \frac{H_{\mathrm{t}, h}}{E_{\mathrm{t}, h}} = \frac{1}{\eta^{(a)}} 
\begin{cases}
k^{(a)} / \beta_h^{(a)}, & \text{TM harmonics}. \\
\beta_h^{(a)} / k^{(a)}, & \text{TE harmonics}.
\end{cases}
\end{align}
where $\beta_{z,h}^{(a)} = \sqrt{\epsilon^{(a)}_r \mu^{(a)}_r k_0^2 - k_{x,m}^2 - k_{y,n}^2}$, $k^{(a)} = \sqrt{\epsilon^{(a)}_r \mu^{(a)}_r k_0}$, and $\eta^{(a)} = \eta_0 \sqrt{\frac{\mu^{(a)}_r}{\epsilon^{(a)}_r}}$ with the impedance of free space, $\eta_0$.

Considering the spatial domain, the surface current excited in the unit cell of the metasurface can be written as
\begin{align}
\mathbf{J}(x, y; \omega) &= F(\omega)\mathbf{J}_p(x, y) \nonumber \\
&= F(\omega) \left \{ \cos \left( \frac{\pi x}{a} \right) \left[ 1 - \left( \frac{2x}{a} \right)^2 \right]^{-1/2} \textrm{rect}{\left( \frac{y}{b} \right)} \hat{\mathbf{x}} \right \}
\label{eq:spatial_current}
\end{align}
This approximation implies that the spatial current distribution $\mathbf{J}_p(x, y)$ of the metasurface is independent of the frequency ($\omega$).
In other words, the frequency dependence can be canceled out in the current distribution of $\mathbf{J}(x, y; \omega)$.
Although it may seem rather restrictive, it has been found accurate for single resonant elements in a wide frequency range, as reported in \cite{rodriguez2015analytical, mesa2018unlocking, mesa2014circuit}.
The approximation is expected to hold up to frequencies below the first high-order resonance of the metasurface in a free-standing case, based on the spatial current distribution under this resonance frequency.

The equivalent circuit network (ECN) analysis of the metasurface, which includes the patch array, is derived by applying the boundary condition at the discontinuity plane ($z = 0$):
\begin{align}
\mathbf{J}_p (x, y) = \hat{\mathbf{z}} \times \left[ \mathbf{H}^{(a)} (x, y) - \mathbf{H}^{(0)} (x, y) \right].
\label{eq:surface_condition}
\end{align}
By substituting the magnetic field expansions from (\ref{eq:magnetic_field1}) and (\ref{eq:magnetic_field2}) into (\ref{eq:surface_condition}) and performing subsequent manipulations, the input impedance ($Z_\mathrm{eq}$) of the ECN model, as shown in Fig. \ref{fig:Equivalent_network_analysis}(c), can be obtained as
\begin{align}
Z_\mathrm{eq} &= \sum_{h \ne 0} \frac{1}{|N_h|^2 \left[Y_h^{(0)}+ Y_h^{(a)}\right]} 
\label{eq:eq_impedance}
\\
N_h & = \tilde{\mathbf{J}}_p^* (\mathbf{k}_{\mathrm{t},h}) \cdot \mathbf{v}_h
\end{align}
where the turn ratio $N_h$ is defined as the coupling between the each harmonic and all the remaining harmonics, and superscript symbol `$\sim$' denotes Fourier transform with respect to $(x, y)$. 

In the case of the finite dielectric/magnetic slab, the Floquet harmonics ($h$) excited on the metasurface propagate inside the dielectric/magnetic substrate with the same harmonic.
For each harmonic, the transmission line with the semi-infinite medium on the right sides is replaced by a corresponding cascade of separated sections, as reported in \cite{garcia2012simplified}.
The admittance $Y_h^{(a)}$ in the equation (\ref{eq:eq_impedance}) need to be replaced by the following impedance $Y_{\mathrm{in},h}^{(\mathrm{d/m})}$:
\begin{align}
Y_{\mathrm{in},h}^{(a)} = Y_{h}^{(a)} \frac{Y_{h}^{(a)} + jY_{h}^{(0)} \tan{(\beta_{z,h}^{(a)}d)}}{Y_{h}^{(0)} + j Y_{h}^{(a)} \tan{(\beta_{z,h}^{(a)}d)}}
\label{eq:input_h}
\end{align}
This equation encompasses the combined effects of both the dielectric/magnetic substrate and the metasurface.

With the ECN model, the reflection coefficient at the discontinuity plane ($z = 0$) is obtained by 
\begin{align}
S_{11} = \frac{Y_0^{(0)} - Y_{\mathrm{in},0} - |N_0|^2 / Z_\mathrm{eq}}
{Y^{(0)}_0 + Y_{\mathrm{in},0} + |N_0|^2 / Z_\mathrm{eq}}.
\label{eq:reflectance}
\end{align}
The excited harmonic is propagating wave with the parallel direction on $\mathbf{k}_{h}^{(0)} = k_{x,m}\hat{\mathbf{x}} + k_{y,n}\hat{\mathbf{y}} + k_{z,h}^{(0)}\hat{\mathbf{z}}$.
The Floquet harmonic of $(m, n) = (0, 0)$ is always the fundamental mode, which forms the incident wave as shown in Fig. \ref{fig:Equivalent_network_analysis}(b).
The variation of the Floquet harmonics with respect to patch dimensions in the free space is analyzed using a patch array with variable $a$ in $P = 3$ mm as shown in Fig. \ref{fig:Equivalent_network_analysis}(d).
This figure demonstrates that the dominant resonance changes with an increase in $a$, although the frequency of perfect impedance matching remains unaffected by changes in $a$. 
In particular, total transmission is observed at these TE cut-off frequencies of the first-order harmonic \cite{garcia2012simplified}.

The unknown coefficient ($N_h$) can be expected from the well-known current profile for specific and canonical geometries \cite{rengarajan2005choice}, such as the rectangular patch.
For the patch array under normal incidence case, this empirical expectation of $N_h$ remains valid up to the maximum effective wavelengths between the \textit{second} resonance and the \textit{third} resonance, i.e., $\lambda \approx  \frac{1 + \sqrt{2}}{2}a$, and the grating lobe regime ($1/P$) \cite{mesa2018unlocking, mesa2014circuit}.
From the perspective of the cavity model, the conventional patch resonator exhibits the cavity resonance modes of TM$_{01}$, TM$_{11}$, and TM$_{20}$ modes, corresponding to the first resonance ($\lambda \approx 2a$), the second resonance ($\lambda \approx \sqrt{2}a$), and the third resonance ($\lambda \approx a$), respectively.
For instance, a metasurface with a maximum dimension of 1.2 mm, permittivity of $\epsilon_r = 6.5 + j0.21$, and 3 mm cell size can be effectively modeled and managed up to 81.2 GHz, corresponding to the second resonance of the metasurface.
Since Floquet harmonics up to this frequency can be engineered, modeling and controlling frequency responses beyond this range presents significant challenges.
For each harmonic, the grounded medium on the right sides can be modeled by a corresponding cascade of separated sections with each shorted transmission line corresponding to the harmonic, as reported in \cite{garcia2012simplified}.
The admittance $Y_{\mathrm{in}, h}^{(a)}$ in the equation (\ref{eq:eq_impedance}) is given by the following equation.
\begin{align}
Y_{\mathrm{in}, h}^{(a)} = -jY_h^{(a)} \cot{(\beta_{z,h}^{(a)} d)}
\label{eq:in_h_short}
\end{align}
This equation encompasses the combined effects of both the dielectric/magnetic substrate and the metasurface. 

}

\bibliographystyle{IEEEtran}
\bibliography{PRX_reference}

\newpage

\vfill

\end{document}